\documentclass[preprints,review,accept,moreauthors,pdftex,10pt,a4paper]{Definitions/mdpi}
\usepackage{graphicx}
\usepackage{dcolumn}
\usepackage{bm}
\usepackage{comment}

\usepackage{environ}
\NewEnviron{myequation1}{%
\begin{equation}
\scalebox{0.96}{$\BODY$}
\end{equation}}
\NewEnviron{myequation2}{%
\begin{equation}
\scalebox{0.96}{$\BODY$}
\end{equation}}
\NewEnviron{myequation3}{%
\begin{equation}
\scalebox{0.96}{$\BODY$}
\end{equation}}
\NewEnviron{myequation4}{%
\begin{equation}
\scalebox{0.94}{$\BODY$}
\end{equation}}

\usepackage[labelformat=simple]{subcaption}

\DeclareCaptionLabelFormat{subcaptionlabel}{\normalfont(\textbf{#2}\normalfont)}
\usepackage{float}
\usepackage{pifont}
\newcommand{\cmark}{\text{\ding{51}}}
\newcommand{\xmark}{\text{\ding{55}}}
\usepackage{caption}
\usepackage{physics}
\usepackage[labelsep=period]{caption}
\usepackage{amsfonts}
\firstpage{1}
\pubvolume{xx}
\issuenum{1}
\articlenumber{5}
\pubyear{2019}
\copyrightyear{2019}
\history{Received: 30 July 2019; Accepted: 29 August 2019; Published: date}
\updates{yes} 

\Title{Directionally-Unbiased Unitary Optical Devices in Discrete-Time Quantum Walks}

\Author{Shuto Osawa $^{1,}$*, David S. Simon $^{1,2,}$* and Alexander V. Sergienko $^{1,3,4,}$*}

\AuthorNames{Firstname Lastname, Firstname Lastname and Firstname Lastname}

\address{%
$^{1}$ \quad Department of Electrical and Computer Engineering, Boston University, 8 Saint Mary’s Street, Boston,
MA~02215, USA\\
$^{2}$ \quad Department of Physics and Astronomy, Stonehill College, 320 Washington Street, Easton, MA
02357, USA\\
$^{3}$ \quad Department of Physics, Boston University, 590 Commonwealth Avenue, Boston, MA 02215,
USA\\
$^{4}$ \quad Photonics Center, Boston University, 8 Saint Mary’s Street, Boston, MA 02215, USA}

\corres{Correspondence: sosawa@bu.edu (S.O.); simond@bu.edu (D.S.S.); alexserg@bu.edu (A.V.S.)}

\abstract{
The optical beam splitter is a widely-used device in photonics-based quantum information processing. Specifically, linear optical networks demand large numbers of beam splitters for unitary matrix realization. This requirement comes from the beam splitter property that a photon cannot go back out of the input ports, which we call ``directionally-biased''. Because of this property, higher dimensional information processing tasks suffer from rapid device resource growth when beam splitters are used in a feed-forward manner. Directionally-unbiased linear-optical devices have been introduced recently to eliminate the directional bias, greatly reducing the numbers of required beam splitters when implementing complicated tasks. Analysis of some originally directional optical devices and basic principles of their conversion into directionally-unbiased systems form the base of this paper. Photonic quantum walk implementations are investigated as a main application of the use of directionally-unbiased systems. Several quantum walk procedures executed on graph networks constructed using directionally-unbiased nodes are discussed. A significant savings in hardware and other required resources when compared with traditional directionally-biased beam-splitter-based optical networks is demonstrated.
}


\begin{document}



\section{Introduction}
The quantum approach to computing attracts public attention mainly because of its capability to execute some computational tasks faster when compared to classical computational devices~\cite{shor1999polynomial, grover1996fast}. Several physical platforms exist to realize quantum computation procedures. Linear optics has been one of the candidates because of its robustness against noise and the ease of quantum state manipulation at room temperature. The design of quantum computing gates with single photons has been proposed and is known as the Knill, Laflamme and Milburn (KLM) 
 model~\cite{knill2001scheme}. This design makes use of linear-optical devices such as beam splitters and phase shifters. The quantum gate performance is executed probabilistically by the process of measuring auxiliary photons. While the KLM model has been used for gate-based quantum computation, other quantum-optical approaches to execute computational tasks have been developed. For example, quantum walks (QW) over optical networks of scattering centers have been considered as another promising tool in executing certain computational tasks~\cite{kempe2003quantum,venegas2012quantum,aharonov1993quantum,portugal2013quantum,childs2009universal}. The construction of such optical networks for quantum walks relies on the use of multiple beam splitters and phase shifters connected in a particular spatial graph pattern. A beam splitter is used as an elementary scattering center during the propagation, and many of them must be cascaded by connecting consecutively in order to form an extensive tree-like network~\cite{carolan2015universal,wang2018multidimensional}. Truly quantum mechanical information processing requires unitarity at every operation. The beam splitter in optics implements two-dimensional unitary transformations and can be seen as a probabilistic mixer of two spatial field modes.

The increase in dimensionality enables employing and manipulating more information, and this needs to be achieved in a coherent way. Optical networks are constructed to perform this task by constructing higher dimensional unitary matrices. It is known that higher dimensional unitary matrices can be decomposed using lower dimensional unitary matrices. By repeating this procedure, any complex unitary matrix can be eventually decomposed using only two-dimensional ones. The Reck decomposition model has been introduced to describe this procedure~\cite{reck1994experimental}. A symmetric version of the Reck model is often called the Clements model~\cite{clements2016optimal}. For instance, these two models have been used by researchers in designing and building experimental linear-optical networks for boson sampling purposes~\cite{spring2013boson,broome2013photonic,tillmann2013experimental,crespi2013integrated}. During the boson sampling process, photons propagate from one side of a complex nodal structure to the other side of the optical network, thus performing a computational task. Direct implementation of multimode optical device has been experimentally verified in integrated platforms~\cite{weihs1996all,peruzzo2011multimode,spagnolo2013three,meany2012non}. Quantum walks over the network of quantum nodes represent another form of quantum information processing, as an alternative to the quantum gate model. QW can also perform certain computations more efficiently than classical algorithms~\cite{aharonov1993quantum,moore2002quantum,krovi2006hitting,childs2002example,childs2003exponential,ambainis2007quantum,magniez2007quantumtri,buhrman2006quantum,magniez2007quantumcom}. Quantum walks in 1D and 2D systems have been experimentally demonstrated in optical systems~\cite{bouwmeester1999optical,knight2003optical,knight2003quantum,schreiber2010photons,pandey2011quantum,zhao2002implement,broome2010discrete,goyal2013implementing,zhang2007demonstration,cardano2015quantum,tang2018experimental,schreiber20122d}.

The traditional quantum walk approach uses a coin operator and a shift operator to execute each elementary step. An alternative description of a quantum walk can be implemented using the scattering quantum walk, also known as the edge walk~\cite{feldman2004scattering,feldman2007modifying}, which has been introduced to describe the quantum walk based on scattering at the nodes or vertices of a lattice on which the walk occurs. There is no need for a coin operator in this model. In order to execute some specific type of quantum walk, we need first to identify a network of scattering centers (a graph) on which the walk is performed. Many different special-purpose graphs can be formed using linear-optical devices in order to execute a particular computational procedure. Thanks to the Reck and Clements decomposition models, the majority of experimental demonstrations in this field, even some complex ones, could be realized using multiple directionally-biased two-dimensional optical devices such as beam splitters. However, the execution of such quantum walks calls for a large number of optical devices when the complexity and the required number of steps in the system increase. This is why quantum walks based on the use of directional devices demand a great deal of costly hardware real estate, which limits their scalability in the long run.

Recently, the original design of a directionally-unbiased linear-optical multiport was introduced~\cite{simon2016group}. This is a unitary coherent optical quantum information processing device that addresses two issues simultaneously: (i) it executes a higher dimensional unitary scattering process at every node of the network with fewer numbers of two-dimensional units for the device construction, and (ii) it scales down significantly the required amount of hardware resources by offering the possibility of reusing scattering units of the graph again and again. An array of such multiports can then form a graph upon which a photon can execute a quantum walk. In principle, the feature of full reversibility can be realized using special designs by incorporating commonly-used optical elements. This is referred to as ``directional'' or ``directionally-biased'' when a photon propagates only in one direction, meaning the input port and the output ports are never the same. This directionality could be circumvented in optics by placing mirrors so that a photon can leave the input port as well. This report will address multiple issues involved in designing, executing, testing, and applying both directional and directionally-unbiased devices. A higher dimensional quantum walk over a graph network based on the use of directionally-unbiased devices will be considered as an example of their practical applications.

\section{Two-Dimensional Linear Optical Devices}

Two-dimensional devices including interferometers are the main building blocks for any applications in classical and in quantum optics. These devices are unitary transformers that mix spatial optical modes without losses and realize the group of 2 $\times$ 2 unitary matrices denoted as $U(2)$. It has been shown that high-dimensional unitary matrices can be decomposed using $U(2)$ matrices~\cite{murnaghan1971unitary}. In order to have flexibility in quantum information processing, one needs to have some means of manipulating amplitude transition coefficients between the input and output fields. In principle, this could be achieved in two ways in optics: (i) by some kind of dynamic change in the input/output splitting ratio of a single beam splitter (BS) or (ii) by forming an interferometer with several beam splitters, thus offering tunability between output ports. In this section, we start with the basic properties of a beam splitter implementing the $U(2)$ operation and discuss its features as a directionally-biased coupler. It will be followed by the consideration of integrated waveguided couplers and some well-known interferometers for implementing 2 $\times$ 2 transformations.

\subsection{Lossless Optical Beam Splitter}

\begin{figure}[H]
\centering
\includegraphics[scale=0.4]{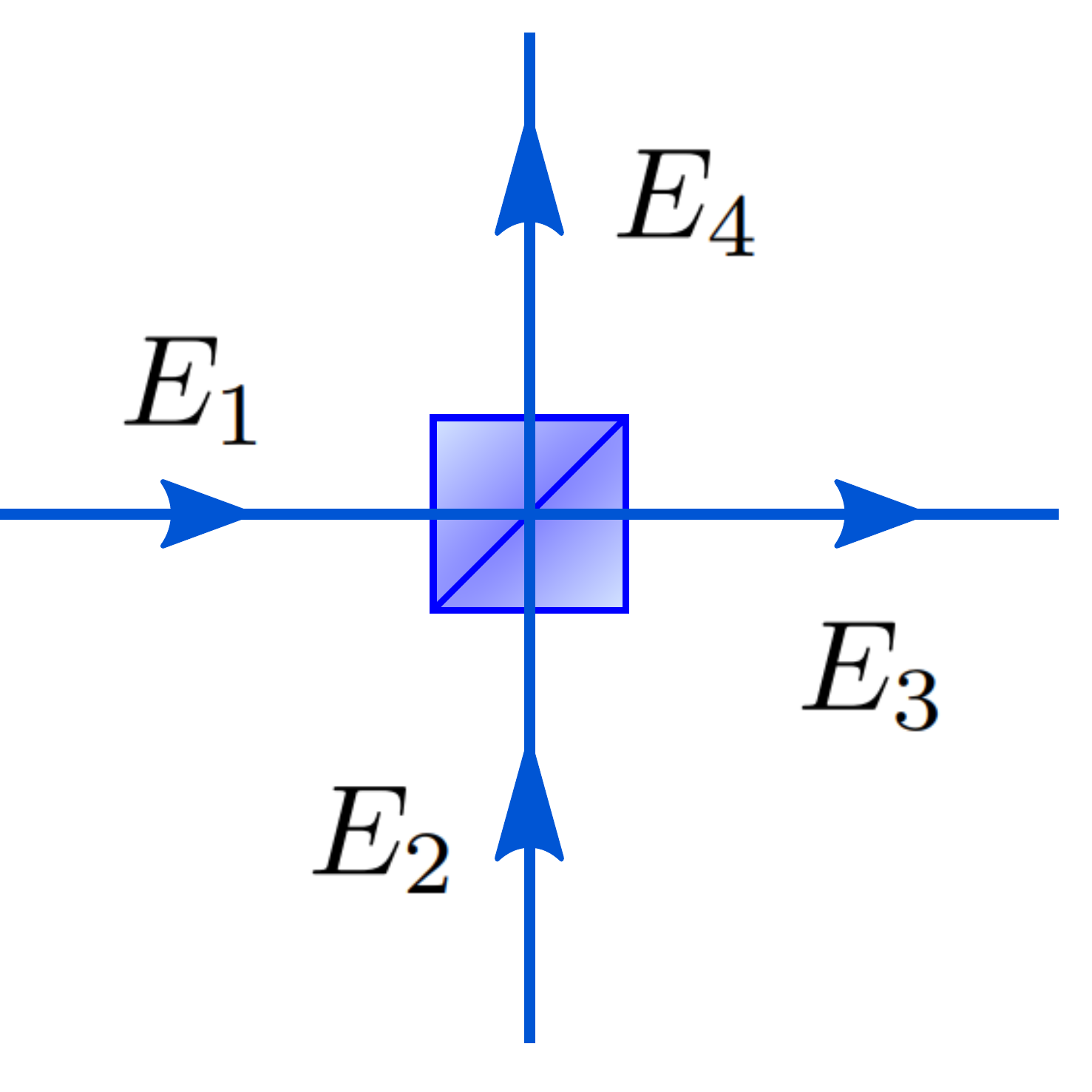}
\caption{Description of a beam splitter. A beam splitter is a device with two input and two output ports. A photon can enter the port $E_1$ and/or $E_2$ and will leave the port $E_3$ and/or $E_4$. \label{fig:BS}}
\end{figure}

A lossless beam splitter introduced in Figure~\ref{fig:BS} redirects incoming photons into two outgoing ports while maintaining energy conservation~\cite{loudon2000quantum,saleh1991fundamentals}. A BS can be represented using a 2 $\times$ 2 matrix, acting on two input and two output ports, denoted as $E_1,E_2$ and $E_3, E_4$, as indicated in Figure~\ref{fig:BS}. The transformation of the fields $E_1,E_2$ is given by:
\begin{equation}
\begin{pmatrix}
E_3\\
E_4
\end{pmatrix}
=
\begin{pmatrix}
T_{13} & R_{23}\\
R_{14} & T_{24}
\end{pmatrix}
\begin{pmatrix}
E_1\\
E_2
\end{pmatrix},
\label{eqn:BS}
\end{equation}

\noindent where $T_{13},T_{24}$ are the transmission from Port 1 to 3 and 2 to 4 and $R_{23},R_{14}$ are reflection for Port 2 to 3 and Port 1 to 4, respectively. The probability conservation relation between the input and output is:
\begin{equation}
|E_3|^2 + |E_4|^2 = |E_1|^2+|E_2|^2.
\label{eqn:energy}
\end{equation}

By substituting Equation~(\ref{eqn:BS}) in Equation~(\ref{eqn:energy}),
\begin{equation}
\begin{split}
|E_3|^2 + |E_4|^2 &= (|T_{13}|^2+|R_{14}|^2)|E_1|^2+(|T_{24}|^2+|R_{23}|^2)|E_2|^2\\
&+T_{13}R_{23}^*E_1E_2^*+R_{23}T_{13}^*E_{2}E_{1}^*+T_{24}R_{14}^*E_2E_1^*+R_{14}T_{24}^*E_{1}E_{2}^*,\\
\end{split}
\end{equation}

\noindent and by comparing the result with Equation~(\ref{eqn:energy}):
\begin{equation}
\begin{split}
|T_{13}|^2+|R_{14}|^2 &= |T_{24}|^2+|R_{23}|^2 = 1\\
T_{13}R_{23}^*+R_{14}T_{24}^*& = R_{23}T_{13}^*+T_{24}R_{14}^* = 0.
\label{eqn:phase}
\end{split}
\end{equation}

Transmission and reflection coefficients \textit{T} and \textit{R} 
 can be rewritten using amplitude and phase. Define $T_{13}=|T_{13}|e^{i\phi_{13}}$, and so on, for all the transmission and reflection coefficients. Then, Equation~(\ref{eqn:phase}) is reduced to:
\begin{equation}
\begin{split}
\frac{|R_{23}|}{|T_{24}|}=-\frac{|R_{14}|}{|T_{13}|}e^{i(\phi_{14}+\phi_{23}-\phi_{24}-\phi_{13})}.
\end{split}
\label{eqn:phaserel1}
\end{equation}

In order to satisfy Equation~(\ref{eqn:phaserel1}), the phase values must be: $\phi_{14}+\phi_{23}-\phi_{24}-\phi_{13} = \pm\pi$.
This phase relation offers some flexibility in choosing the phase settings. Two different phase settings often appear in the literature for a beam splitter with a 50/50 power splitting ratio. When $|R_{23}|=|R_{14}|=|R|=\frac{1}{\sqrt{2}},|T_{13}|=|T_{24}|=|T|=\frac{1}{\sqrt{2}}$, one could choose $\phi_{14}=\phi_{13}=\phi_{23} = 0,\phi_{24}=\pi$ as an example. Other splitting ratios can be chosen as long as $|T|^2+|R|^2=1$ is satisfied.

\vspace{6pt}
\noindent Example 1:
\begin{equation}
BS_{1} =
\frac{1}{\sqrt{2}}
\begin{pmatrix}
1 & 1\\
1 & -1
\end{pmatrix}.
\end{equation}

Other phase settings could be $\phi_{23}=\phi_{14}=\phi_{R},\phi_{13}=\phi_{24}=\phi_{T}$ with $|R_{23}|=|R_{14}|=|R|,|T_{13}|=|T_{24}|=|T|$ and substituting these in Equation~(\ref{eqn:phaserel1}).

\begin{equation}
\begin{split}
\frac{|R|}{|T|}=-\frac{|R|}{|T|}e^{2i(\phi_{T}-\phi_{R})},
\end{split}
\end{equation}

\noindent where $\phi_{R}-\phi_{T}=\frac{\pi}{2}$.

By choosing the phase settings $\phi_T=0, \phi_R = \frac{\pi}{2}$, another example of the BS matrix can be produced.

\vspace{6pt}
\noindent Example 2:
\begin{equation}
BS_{2} =
\frac{1}{\sqrt{2}}
\begin{pmatrix}
1 & i\\
i & 1
\end{pmatrix}.
\end{equation}

Both examples are equivalent when appropriate phase shifters have been introduced before and after the beam splitter:
\begin{equation}
\frac{1}{\sqrt{2}}
\begin{pmatrix}
1 & 1\\
1 & -1
\end{pmatrix}
=
\begin{pmatrix}
1 & 0\\
0 & e^{-i\frac{\pi}{2}}
\end{pmatrix}
\frac{1}{\sqrt{2}}
\begin{pmatrix}
1 & i\\
i & 1
\end{pmatrix}
\begin{pmatrix}
1 & 0\\
0 & e^{-i\frac{\pi}{2}}
\end{pmatrix}.
\end{equation}

\subsection{Directionality of a Beam Splitter}

A beam splitter is a symmetric device, meaning that any one of four ports can be used as an input, and its action is invariant under time reversal. At the same time, the device is not symmetric in the sense that the incoming photon cannot leave through the input port. We call this feature ``directional-bias''; the choice of an input port biases the output to be in only two of the four possible output directions. This directional bias increases the required number of beam splitters when one enters the realm of higher dimensionality. In principle, this directional bias could be circumvented by placing external mirrors after the beam splitters so that they reverse the light propagation direction. This would allow the photon to leave through the input ports, and the system now becomes ``directionally-unbiased''; all four output possibilities can still be realized, regardless of an input direction. Examples of ways to achieve this will be discussed in the coming sections.

\subsection{2 $\times$ 2 Integrated Directional Waveguide Coupler}

\begin{figure}[H]
 \centering
 \begin{subfigure}[t]{0.5\textwidth}
 \centering
 \includegraphics[height=1.2in]{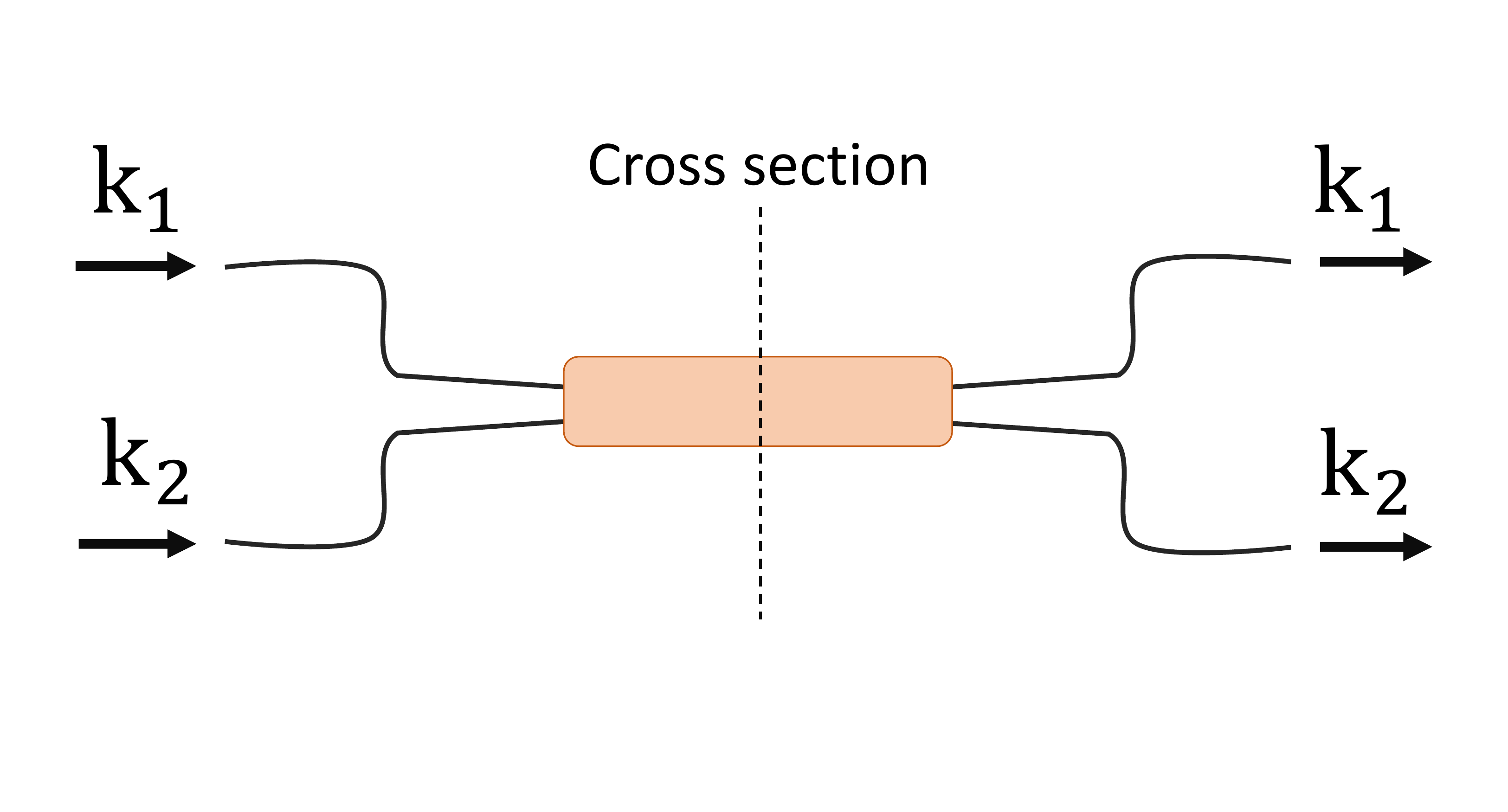}
 \caption{}\label{fig:glued_tree}
 \end{subfigure}%
 ~
 \begin{subfigure}[t]{0.5\textwidth}
 \centering
 \includegraphics[height=1.2in]{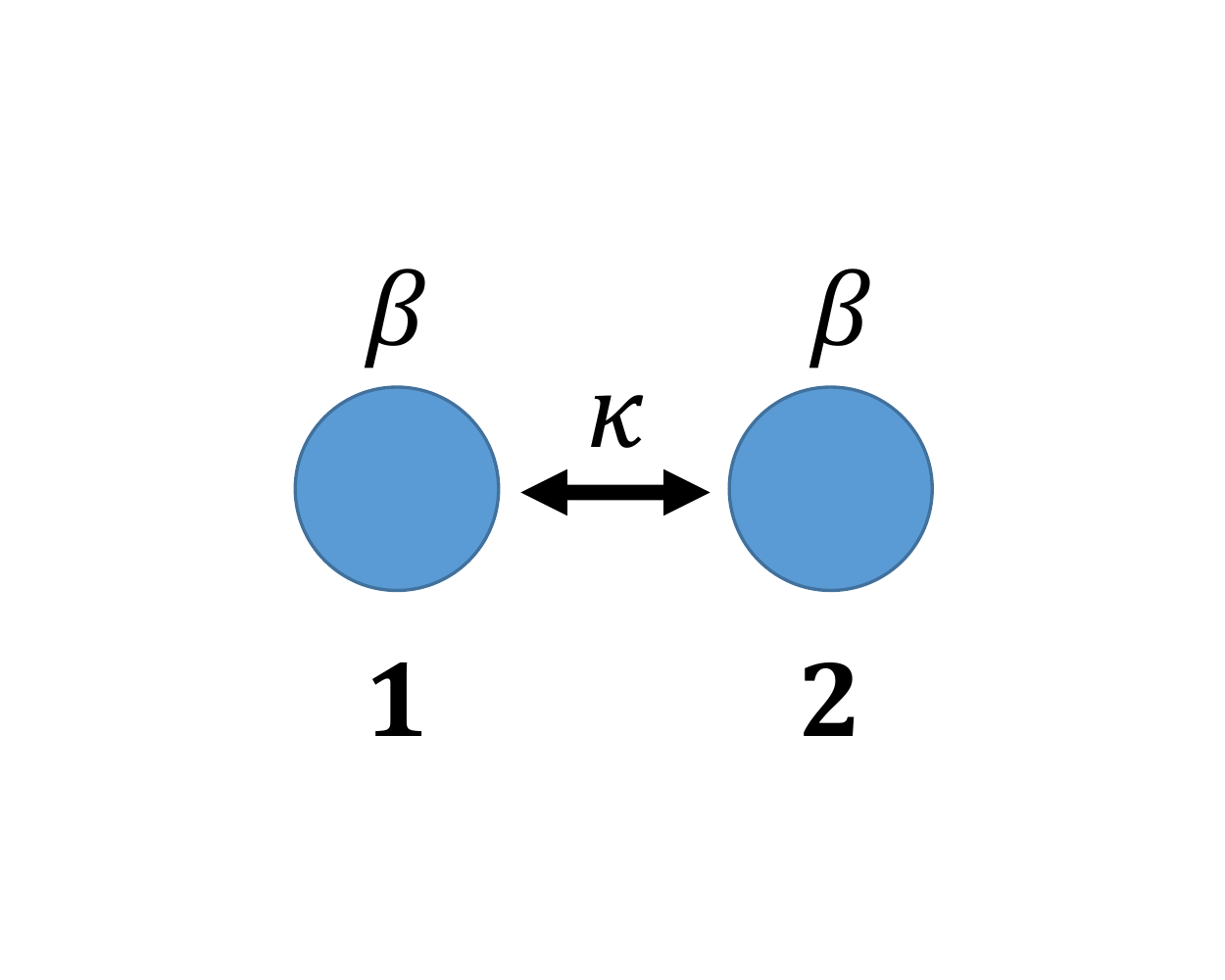}
 \caption{}\label{fig:4d_hyper}
 \end{subfigure}
 \caption{(\textbf{a}) A 2 $\times$ 2 directional coupler. $k_1$ and $k_2$ represent input and output spatial modes. A photon can enter either $k_1$ or $k_2$, and a coupler transforms the input state. Two waveguides are closely located to allow evanescent coupling at the cross-section in the figure. (\textbf{b}) The cross-section area in the directional coupler illustrated in (\textbf{a}). The coupling strength between two waveguides is $\kappa$, and the propagation constant in each waveguide is~$\beta$. \label{fig:integrated_2x2}}
\end{figure}

A directional coupler is an integrated optics analog of a beam splitter. When two waveguides are brought close together, evanescent waves overlap and start coupling in the neighboring waveguide. Figure~\ref{fig:integrated_2x2} illustrates a 2 $\times$ 2 integrated directional coupler and its cross-section. The coupling strength $\kappa$ can be controlled by changing the distance between two waveguides. The BS and a directional coupler are both directionally-biased devices. The propagation of a photon through these devices can be described using a transfer matrix U. $E_{out} = UE_{in}$, where $E_{in}$ and $E_{out}$ are the input and output~fields.

The transfer matrix of the directional coupler can be derived using coupled mode equations based on the Heisenberg equation~\cite{bromberg2009quantum,spagnolo2013three}. The evolution in the $z$ direction is given by:
\begin{equation}
\begin{split}
 i\frac{dA_1^\dagger}{dz} = \beta A_1^\dagger + \kappa A_2^\dagger, \\
 i\frac{dA_2^\dagger}{dz} = \kappa A_1^\dagger + \beta A_2^\dagger,\\
 \end{split}
 \label{eqn:coupling2x2}
\end{equation}

\noindent where $A_j^\dagger$ $\{j = 1,2\}$ are creation operators for a photon in the $j^{\text{th}}$ waveguide. $\beta$ is a waveguide propagation constant, and $\kappa$ is a coupling coefficient between the two waveguides. \\
Equation~(\ref{eqn:coupling2x2}) can be rewritten in a matrix form:
\begin{equation}
\begin{pmatrix}
\frac{dA_1^\dagger}{dz}\\
\frac{dA_2^\dagger}{dz}
\end{pmatrix}
=-i
\begin{pmatrix}
\beta & \kappa \\
\kappa & \beta\\
\end{pmatrix}
\begin{pmatrix}
A_1\\
A_2
\end{pmatrix}.
\label{eqn:matrix_2x2}
\end{equation}

We can solve for $A_1$ and $A_2$ by Equation~(\ref{eqn:coupling2x2}) using differential equation solutions in the form of Equation~(\ref{eqn:matrix_2x2}) and finding eigenvalues and eigenvectors.

Eigenvalues with corresponding eigenvectors are given by:
\begin{equation*}
\lambda_1 = -\beta i-\kappa i:
\begin{pmatrix}
1\\
1
\end{pmatrix},
\lambda_2 = -\beta i+\kappa i:
\begin{pmatrix}
-1\\
1
\end{pmatrix},
\end{equation*}
\begin{equation}
\begin{pmatrix}
A_1\\
A_2
\end{pmatrix}
=
c_1e^{-(\beta z +\kappa z)i}
\begin{pmatrix}
1\\
1
\end{pmatrix}
+c_2e^{-(\beta z-\kappa z)i}
\begin{pmatrix}
-1\\
1
\end{pmatrix},
\end{equation}

Initial conditions are given by: $A_1(0) = 1,A_2(0) = 0$, and $c_1 = \frac{1}{2}, c_2 = -\frac{1}{2}.$ The full transfer matrix can be reconstructed after solving also for the alternative initial condition: $A_1(0) = 0, A_2(0) = 1$.
\begin{equation}
U_{Coupler} = \frac{e^{-\beta z i}}{2}
\begin{pmatrix}
e^{-\kappa z i}+e^{\kappa z i} &e^{-\kappa z i}-e^{\kappa z i}\\
e^{-\kappa z i}-e^{\kappa z i}&e^{-\kappa z i}+e^{\kappa z i}
\end{pmatrix}
=e^{-\beta z i}
\begin{pmatrix}
cos(\kappa z) & -isin(\kappa z)\\
-isin(\kappa z)&cos(\kappa z)
\end{pmatrix}.
\end{equation}

\subsection{Interferometers as Two-Dimensional Devices}

Interferometers are essential tools in quantum information processing and usually involve multiple beam splitters. The amplitude of each of the two outgoing modes can be modified by changing the relative phase between two paths. There are several major interferometer designs that offer 2 $\times$ 2 mode transformation. The Mach--Zehnder interferometer is a directionally-biased device that could be useful in realizing the Reck decomposition model, while the Michelson interferometer does not suffer from directional bias.

\subsubsection{Mach--Zehnder Interferometer}

\begin{figure}[H]
\includegraphics[scale=0.3]{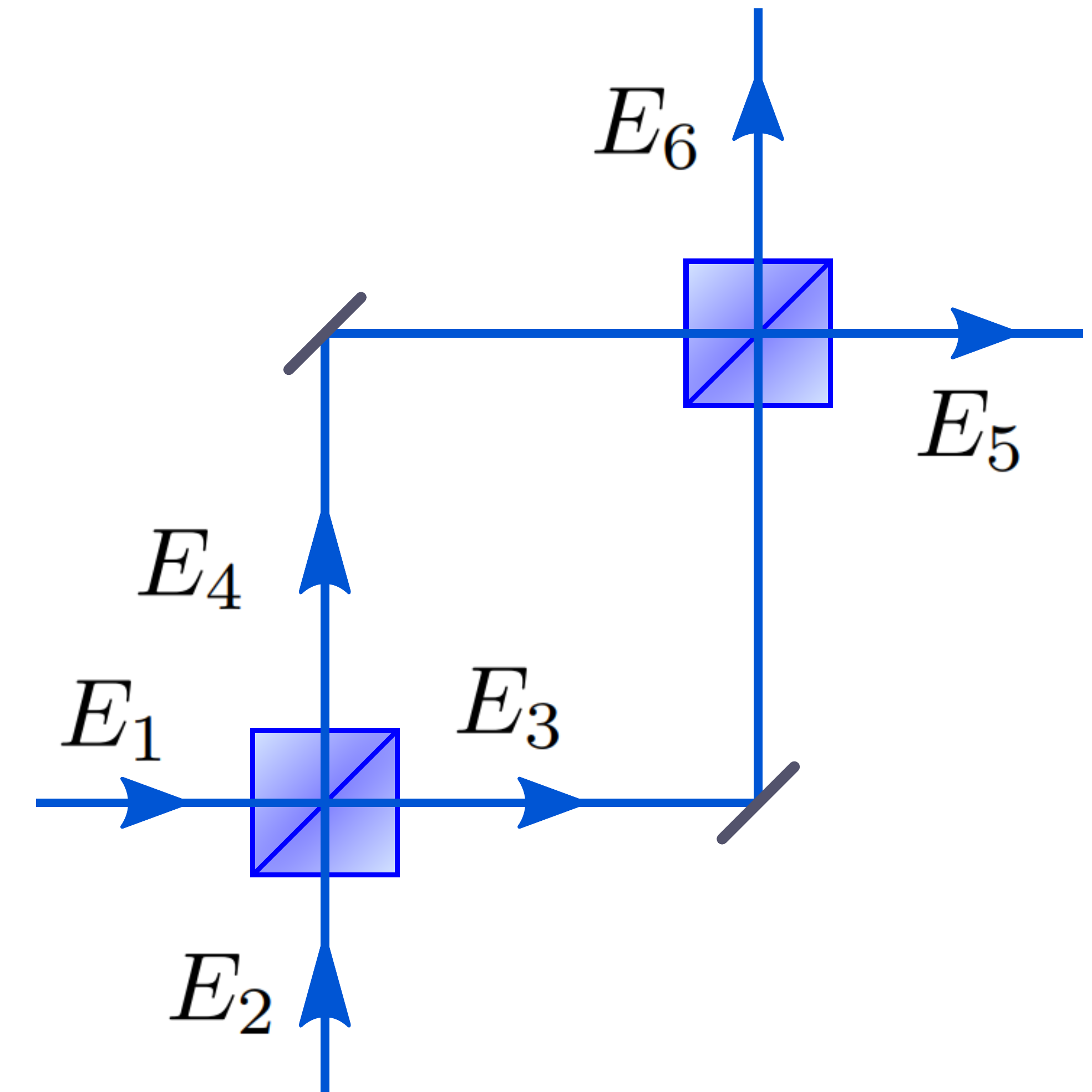}
\centering
\caption{The Mach--Zehnder interferometer. A photon can enter either the port $E_1$ or $E_2$ and will be transformed by the first beam splitter. The photon can leave either through a superposition of $E_3$ and $E_4$ and be transformed by the second beam splitter. Finally, the photon leaves the device either through $E_5$ and/or $E_6$ modes.}
\label{fig:2x2_MZ}
\end{figure}

The Mach--Zehnder interferometer shown in Figure~\ref{fig:2x2_MZ} is a {directionally-biased} interferometer. Assume that each beam splitter has a 50/50 power splitting ratio between the two outgoing fields and the device is symmetric because the path length between two arms can be made identical. The beam splitter matrix $U_{BS}$ is applied twice, and the relative phase shift $\phi$ between the two modes by applying the matrix $U_{phase}$ is introduced before the photon encounters the second beam splitter.
\begin{equation}
U_{MZ} = U_{BS}U_{phase}U_{BS}.
\end{equation}

By introducing expressions for the above matrices, one would obtain a specific formulation of the Mach--Zehnder transformation:
\begin{equation}
\begin{split}
U_{MZ} &=
\frac{1}{\sqrt{2}}
\begin{pmatrix}
1 & i\\
i & 1
\end{pmatrix}
\begin{pmatrix}
e^{i\phi} & 0\\
0 & 1
\end{pmatrix}
\frac{1}{\sqrt{2}}
\begin{pmatrix}
1 & i\\
i & 1
\end{pmatrix}
=\frac{1}{2}
\begin{pmatrix}
e^{i\phi}-1 & i(e^{i\phi}+1)\\
 i(e^{i\phi}+1) &1-e^{i\phi}
\end{pmatrix}\\
&=\frac{1}{2}
\begin{pmatrix}
e^{i\frac{\phi}{2}}(e^{i\frac{\phi}{2}}-e^{-i\frac{\phi}{2}}) & ie^{i\frac{\phi}{2}}(e^{i\frac{\phi}{2}}+e^{-i\frac{\phi}{2}})\\
 ie^{i\frac{\phi}{2}}(e^{i\frac{\phi}{2}}+e^{-i\frac{\phi}{2}}) & -ie^{i\frac{\phi}{2}}(e^{i\frac{\phi}{2}}-e^{-i\frac{\phi}{2}})
\end{pmatrix}
=e^{i(\frac{\phi}{2}+\frac{\pi}{2})}
\begin{pmatrix}
sin(\frac{\phi}{2}) & cos(\frac{\phi}{2})\\
cos(\frac{\phi}{2}) & -sin(\frac{\phi}{2})
\end{pmatrix}.
\end{split}
\end{equation}

The element-wise multiplication leads to the following input-output probability distribution:
\begin{equation}
P_{MZ}=U_{MZ}U_{MZ}^*=
\frac{1}{2}
\begin{pmatrix}
1-cos{\phi} &1+cos{\phi}\\
1+cos{\phi} &1-cos{\phi}
\end{pmatrix},
\end{equation}

\noindent where $U^*$ is a complex conjugate of $U$.

One can easily see that the Mach--Zehnder interferometer effectively serves as a {tunable directionally-biased variable beam splitter}, and this tunability plays a key role in higher dimensional interferometer-based optical networks.

\subsubsection{Michelson Interferometer}

\begin{figure}[H]
\centering
\includegraphics[scale=0.3]{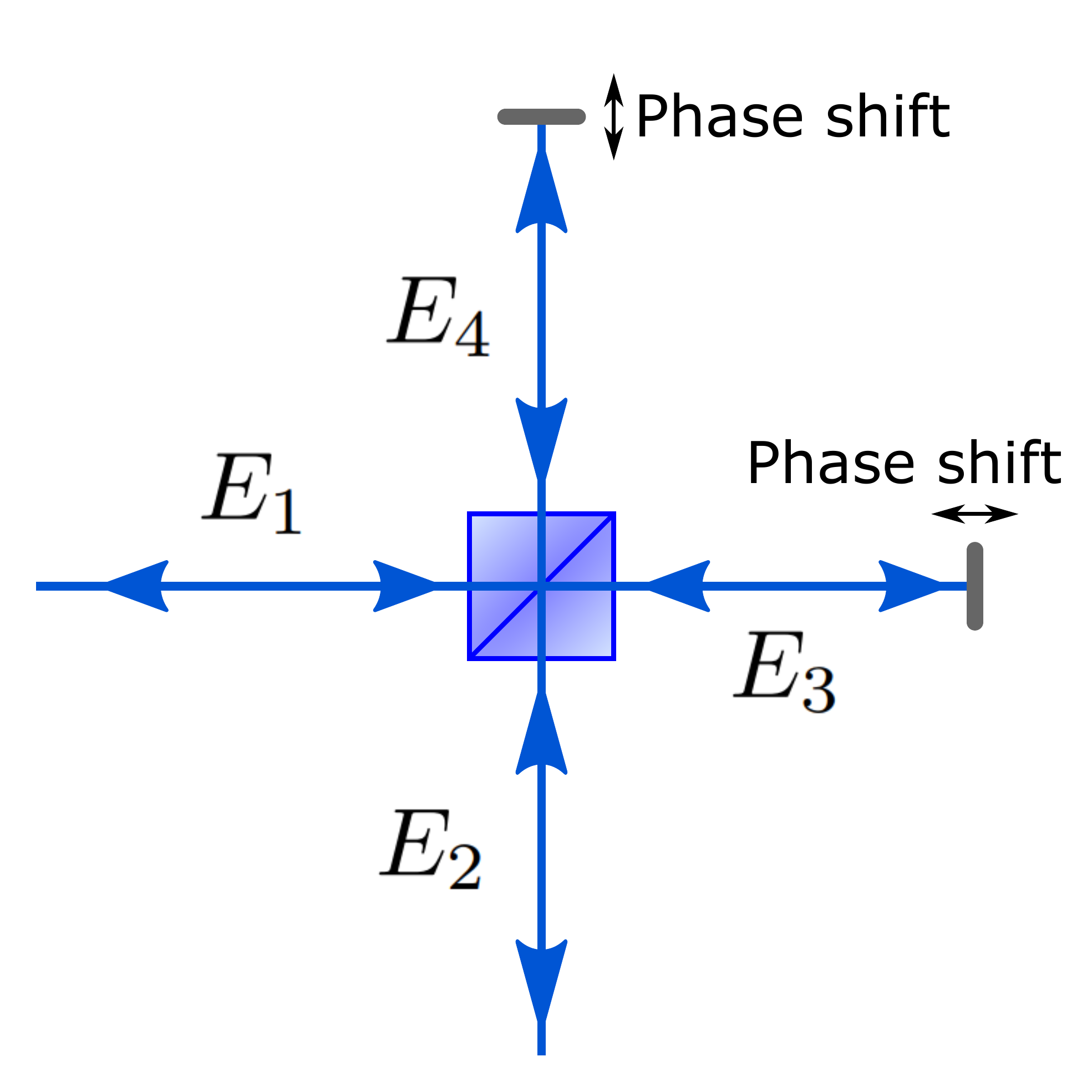}
\caption{The Michelson interferometer. A photon can enter either the port $E_1$ and/or $E_2$. The photon interacts with a beam splitter, mirror, and again with a beamsplitter. The photon leaves through either $E_1$ and/or $E_2$, which are the same as the input ports. The phase can be controlled by translating mirrors in the system. \label{fig:2x2_michelson}}
\end{figure}

The Michelson interferometer in Figure~\ref{fig:2x2_michelson} is one example of the {directionally-unbiased} 2 $\times$ 2 device. Its layout could be used as an illustration of a general optical design principle that the directional bias within a 2 $\times$ 2 device could be circumvented by placing mirrors after the first beam splitter encounter and reversing directions of the optical flux.

A state of the input photon will be transformed by the first interaction with a BS.
\begin{equation}
\begin{pmatrix}
E_3\\
E_4
\end{pmatrix}
=
U_{BS}
\begin{pmatrix}
E_1\\
E_2
\end{pmatrix},
\end{equation}

\noindent where $U_{BS}$ represents a linear transformation between the input fields and output fields of a beam splitter. In order to consider the transformation of fields from $E_3$ and $E_4$ to $E_1$ and $E_2$, one must multiply the outcome with the inverse of matrix $U_{BS}$ from the left.
\begin{equation}
U_{BS}^{-1}
\begin{pmatrix}
E_3\\
E_4
\end{pmatrix}
=
U_{BS}^{-1}
U_{BS}
\begin{pmatrix}
E_1\\
E_2
\end{pmatrix}.
\label{eqn:michelson}
\end{equation}

Since $U_{BS}$ is a unitary matrix $U_{BS}^{-1} = U_{BS}^{\dagger}$, then:

\begin{equation}
\begin{pmatrix}
E_1\\
E_2
\end{pmatrix}
=
U_{BS}^{\dagger}
\begin{pmatrix}
E_3\\
E_4
\end{pmatrix}.
\end{equation}

This describes the transformation from $E_3$ and $E_4$ to $E_1$ and $E_2$. This still represents a forward propagation; therefore, we must take a complex conjugate to reverse the propagation direction.
\begin{equation}
\begin{pmatrix}
E_1^*\\
E_2^*
\end{pmatrix}
=
U_{BS}^{T}
\begin{pmatrix}
E_3^*\\
E_4^*
\end{pmatrix}.
\end{equation}

This equation is read as a reverse propagation from $E_3$ and $E_4$ to $E_1$ and $E_2$. Now, the 2 $\times$ 2 transformation of optical modes by the Michelson interferometer is described as:
\begin{equation}
U_{Michelson} = U_{BS}^{T}U_{Phase}U_{BS}.
\end{equation}

The first transformation $U_{BS}$ describes the propagation from $E_1$ and $E_2$ to $E_3$ and $E_4$, and then, phase shift $U_{phase}$ introduces phase shifts between the two fields. The phases can be controlled by translating mirrors in the system as indicated in Figure~\ref{fig:2x2_michelson}. Finally, the reversed propagation and transformation from $E_3$ and $E_4$ to $E_1$ and $E_2$ is given by $U_{BS}^{T}$. These transformations complete the transformation of input fields by the Michelson interferometer. Such 2 $\times$ 2 directionally-unbiased devices based on the Michelson interferometer configuration could be used as elements for building higher dimensional interferometric systems. The Michelson interferometer is essentially the two-port version of the unbiased multiports introduced below, with $U_{phase}$ providing the tunability.

\section{Three- and Four-Dimensional Linear Optical Devices}
The dimensionality of the 2 $\times$ 2 linear-optical device investigated in the previous section can be expanded to a more general situation covering a greater number of spatial modes. It has been shown in the past that one has to rely on using multiple 2 $\times$ 2 beam splitters in order to execute a high-dimensional transformation. This relationship is often called a Reck decomposition model (Reck model)~\cite{reck1994experimental}. The Reck model has been demonstrated experimentally~\cite{carolan2015universal}. There is also a symmetric directional alternative to Reck's approach that is called Clements' design~\cite{sansoni2012two,clements2016optimal}. This design can realize any unitary matrices and will be discussed in the 4 $\times$ 4 device section. In addition to Reck's and Clements' decomposition via multiple lower dimensional devices, a 3 $\times$ 3 directional transformation could be realized directly by exploiting a 3D optical integrated device in a waveguide configuration that is called an optical tritter~\cite{kowalevicz2005three,suzuki2006characterization,meany2012non}. Another decomposition model has been proposed as well~\cite{de2018simple}. This section examines these possible designs for 3 $\times$ 3 and 4 $\times$ 4 devices in detail. Four-dimensional devices are not just a simple extension of the three-dimensional devices. When the numbers of ports exceeds three, the distances between couplers are not identical. This means coupling strength would not be the same between couplers; therefore, it can change the final transfer matrix between the input fields and the output fields. 

\subsection{Reck Decomposition Design}

It has been shown theoretically~\cite{reck1994experimental} that an arbitrary single N $\times$ N unitary matrix, U(N), can be decomposed into a succession of $\frac{N(N-1)}{2}$ numbers of 2 $\times$ 2 mode mixing matrices. In order to understand the decomposition procedure, it is useful to understand the decomposition procedure for the 2 $\times$ 2 unitary matrix. Higher dimensional decomposition examples will be provided after the 2~$\times$~2 example.
An arbitrary unitary 2 $\times$ 2 matrix U(2) is defined as:
\begin{equation}
U(2) =
\begin{pmatrix}
A & B\\
C& D
\end{pmatrix},
\end{equation}

\noindent where $ A, B, C, D \in \mathbb{C}$. $\mathbb{C}$ is a set of complex numbers. It is always possible to find a unitary matrix T such that U(2) becomes diagonal after it is multiplied by the matrix T.
\begin{equation}
U(2)T =
\begin{pmatrix}
A' & 0\\
0& D'
\end{pmatrix},
\end{equation}
\noindent where $ A', D' \in \mathbb{C}$.

The resulting diagonalized matrix will be turned into an identity matrix by multiplying it with an additional diagonal matrix P.
\begin{equation}
U(2)TP =
\begin{pmatrix}
1 & 0\\
0& 1
\end{pmatrix}.
\end{equation}

This procedure shows that any U(2) matrix can be transformed into an identity matrix. This result indicates that the inverse matrix $(TP)^{-1}$ is the original U(2) we wanted. T and P are both unitary matrices; therefore, $T^{-1} = T^{\dagger}$ and $P^{-1} = P^{\dagger}$ where $\dagger$ is complex conjugate and transpose.
\begin{equation}
\begin{split}
U(2)TP = I(2) \rightarrow U(2) = (TP)^{-1}=P^\dagger T^\dagger.\\
\end{split}
\end{equation}

This procedure shows that a matrix U(2) is decomposed into matrices $P^\dagger$ and $T^\dagger$. Such a diagonalization process can be applied in higher dimensions as well. In the 3 $\times$ 3 case, arbitrary U(3) matrices can be diagonalized using multiple 3 $\times$ 3 matrices with each matrix containing U(2) inside. $T_{3,1}$, $T_{2,1}$, and $T_{3,2}$ are the matrices containing U(2) inside. $T_{3,1}$ mixes spatial Modes 3 and 1; $T_{2,1}$~mixes spatial Modes 2 and 1; and $T_{3,2}$ mixes spatial Modes 3 and 2.

\begin{equation}
T_{3,1} =
\begin{pmatrix}
U(2)_{1,1} & 0 & U(2)_{1,2}\\
0 & 1 & 0\\
U(2)_{2,1} & 0 & U(2)_{2,2}
\end{pmatrix},
T_{3,2} =
\begin{pmatrix}
1 & 0\\
0 & U(2)
\end{pmatrix},
T_{2,1} =
\begin{pmatrix}
U(2) & 0\\
0 & 1
\end{pmatrix},
\end{equation}
\noindent where $U(2)_{1,1}$ is the element of U(2) from the first row and the first column. The rest of three elements $U(2)_{1,2},U(2)_{2,1},U(2)_{2,2}$ follow the same rule. Our goal is to find a decomposition for an arbitrary matrix U(3). Assume that the U(3) matrix has the form of:
\begin{equation}
U(3) =
\begin{pmatrix}
A _1& B_1&C_1\\
D_1& E_1&F_1\\
G_1&H_1&I_1
\end{pmatrix}.
\end{equation}

As a first step, elements from the first row and the third column $U(3)_{1,3}$ and the third row and the first column $U(3)_{3,1}$ can be eliminated by multiplying a matrix $T_{3,2}$.
\begin{equation}
U(3)T_{3,2}=
\begin{pmatrix}
A_2 & B_2&0\\
D_2& E_2&F_2\\
0&H_2&I_2
\end{pmatrix}.
\end{equation}

Repeat the elimination procedure for all the non-diagonal elements of U(3).
\begin{equation}
U(3)T_{3,2}T_{3,1}=
\begin{pmatrix}
A_3 & B_3&0\\
D_3& E_3&0\\
0&0&I_3
\end{pmatrix},
\end{equation}
\begin{equation}
U(3)T_{3,2}T_{3,1}T_{2,1}=
\begin{pmatrix}
A_4 & 0&0\\
0& E_4&0\\
0&0&I_4
\end{pmatrix}.
\end{equation}

U(3) is transformed into an identity matrix after multiplying by a diagonal matrix P. U(3) can be obtained by taking the inverse of $(T_{3,2}T_{3,1}T_{2,1}P)$.
\begin{equation}
\begin{split}
U(3)T_{3,2}T_{3,1}T_{2,1}P &= I(3) \rightarrow U(3) = (T_{3,2}T_{3,1}T_{2,1}P)^{-1} =P^\dagger T_{2,1}^\dagger T_{3,1}^\dagger T_{3,2}^\dagger.\\
\end{split}
\end{equation}

Matrix elements $A_i$ through $I_i$, $ i \in \mathbb{Z}$ belong to $\mathbb{C}$. $\mathbb{Z}$ is a set of integers. The U(3) matrix is decomposed into matrices $P^\dagger$, $T_{2,1}^\dagger$, $T_{3,1}^\dagger$, and $T_{3,2}^\dagger$. This concludes that the knowledge of each individual beam splitter (or interferometer) in the system allows reconstructing a transfer matrix of the whole system. The same reconstruction process can be applied in the 4 $\times$ 4 case of U(4) decomposition.
\begin{equation}
\begin{split}
U(4)T_{4,3}T_{4,2}T_{4,1}T_{3,2}T_{3,1}T_{2,1}P &= I(4)\\ 
 \rightarrow U(4) = (T_{4,3}T_{4,2}T_{4,1}T_{3,2}T_{3,1}T_{2,1}P)^{-1} &= P^\dagger T_{2,1}^\dagger T_{3,1}^\dagger T_{3,2}^\dagger T_{4,1}^\dagger T_{4,2}^\dagger T_{4,3}^\dagger.
\end{split}
\end{equation}

\begin{figure}[H]
 \centering
 \begin{subfigure}[t]{0.5\textwidth}
 \centering
 \includegraphics[height=1.8in]{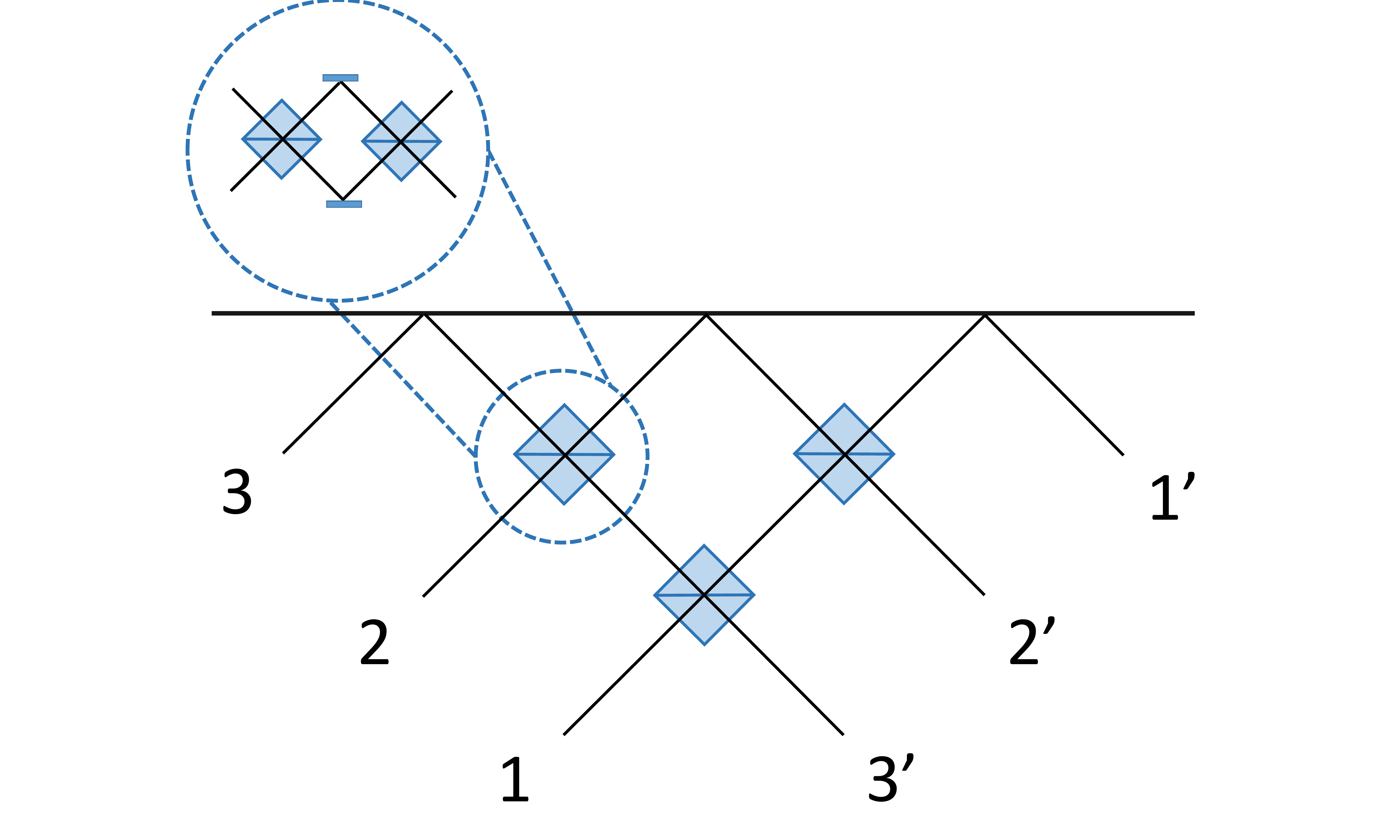}
 \caption{}\label{fig:glued_tree}
 \end{subfigure}%
 ~
 \begin{subfigure}[t]{0.5\textwidth}
 \centering
 \includegraphics[height=1.8in]{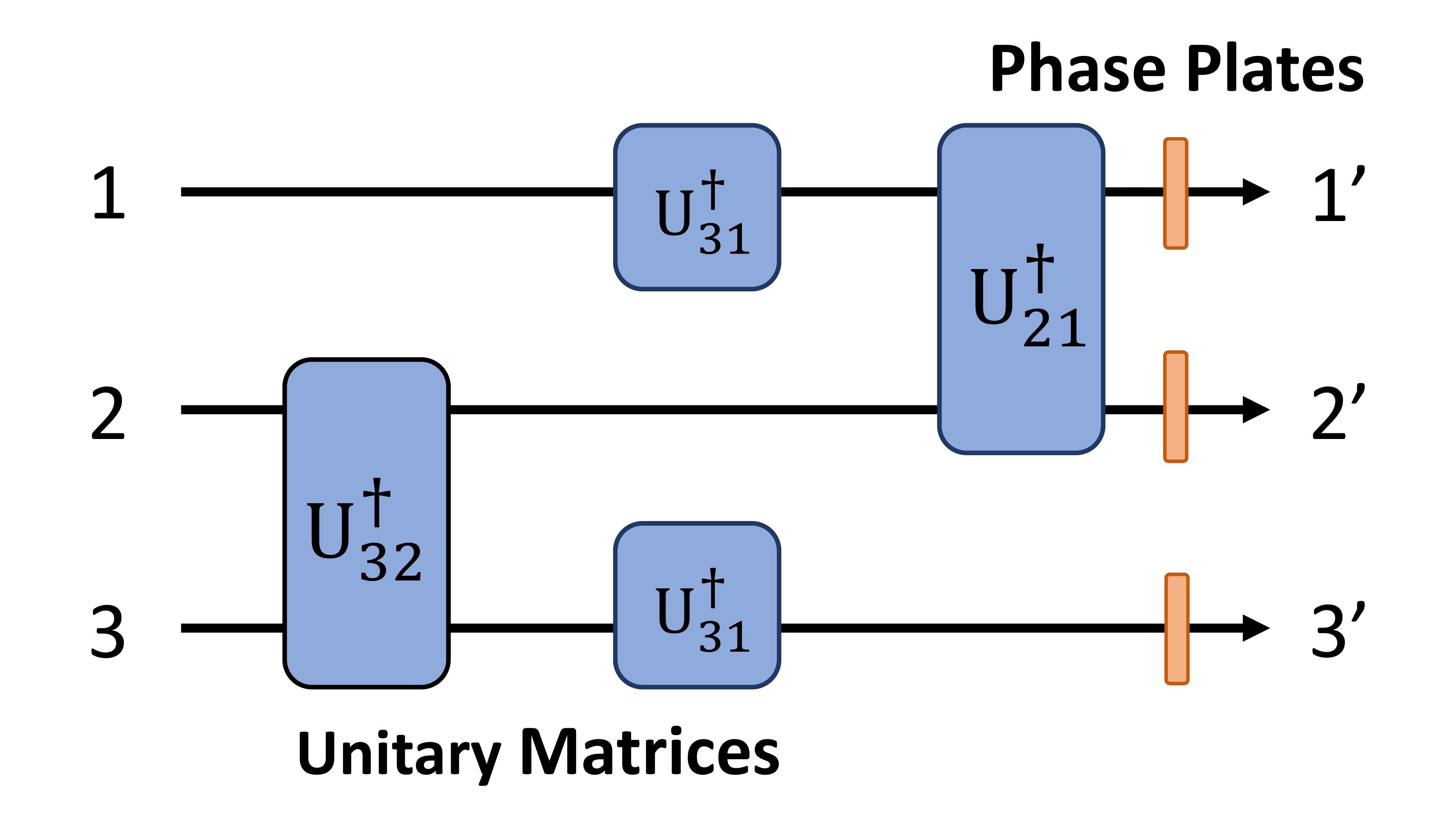}
 \caption{}\label{fig:4d_hyper}
 \end{subfigure}
 \caption{(\textbf{a}) The 3 $\times$ 3 Reck model realization using three beam splitters. A photon can enter either Port 1, 2, or 3 and the photon leave either Port 1', 2', and/or 3'. A beam splitter in the system can be substituted by a Mach--Zehnder interferometer if one wants to give amplitude tuning at each beam splitter encounter. The beam splitter requirement will be increased to six when amplitude tuning by interferometers is imposed. (\textbf{b}) The information flow decomposition of the bulk 3 $\times$ 3 setup using a set of 2 $\times$ 2 unitary matrices. This is equivalent to the physical setup in (\textbf{a}). \label{fig:reck_three_bulk}}
\end{figure}

\begin{figure}[H]
 \centering
 \begin{subfigure}[t]{0.5\textwidth}
 \centering
 \includegraphics[height=1.9in]{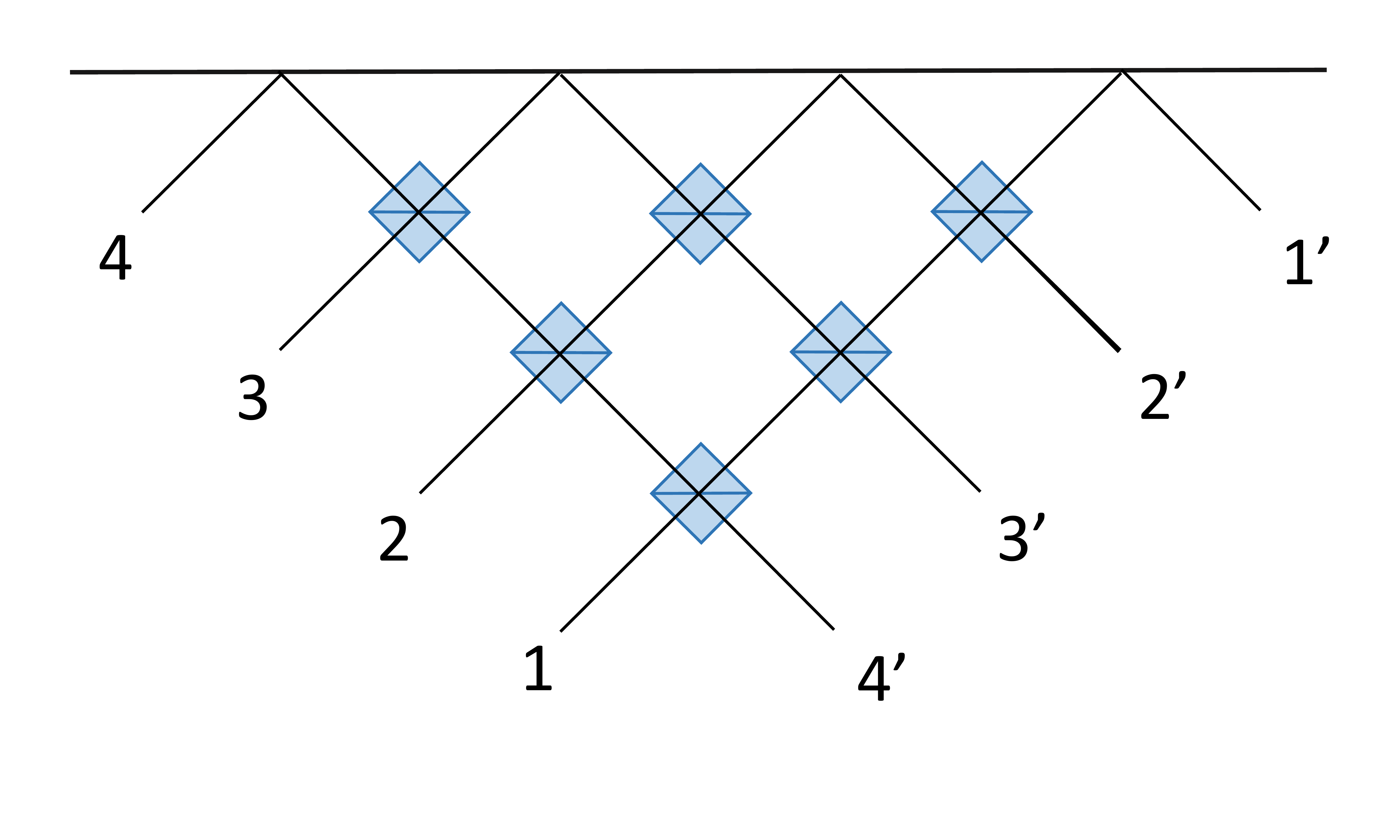}
 \caption{}\label{fig:glued_tree}
 \end{subfigure}%
 ~
 \begin{subfigure}[t]{0.5\textwidth}
 \centering
 \includegraphics[height=1.9in]{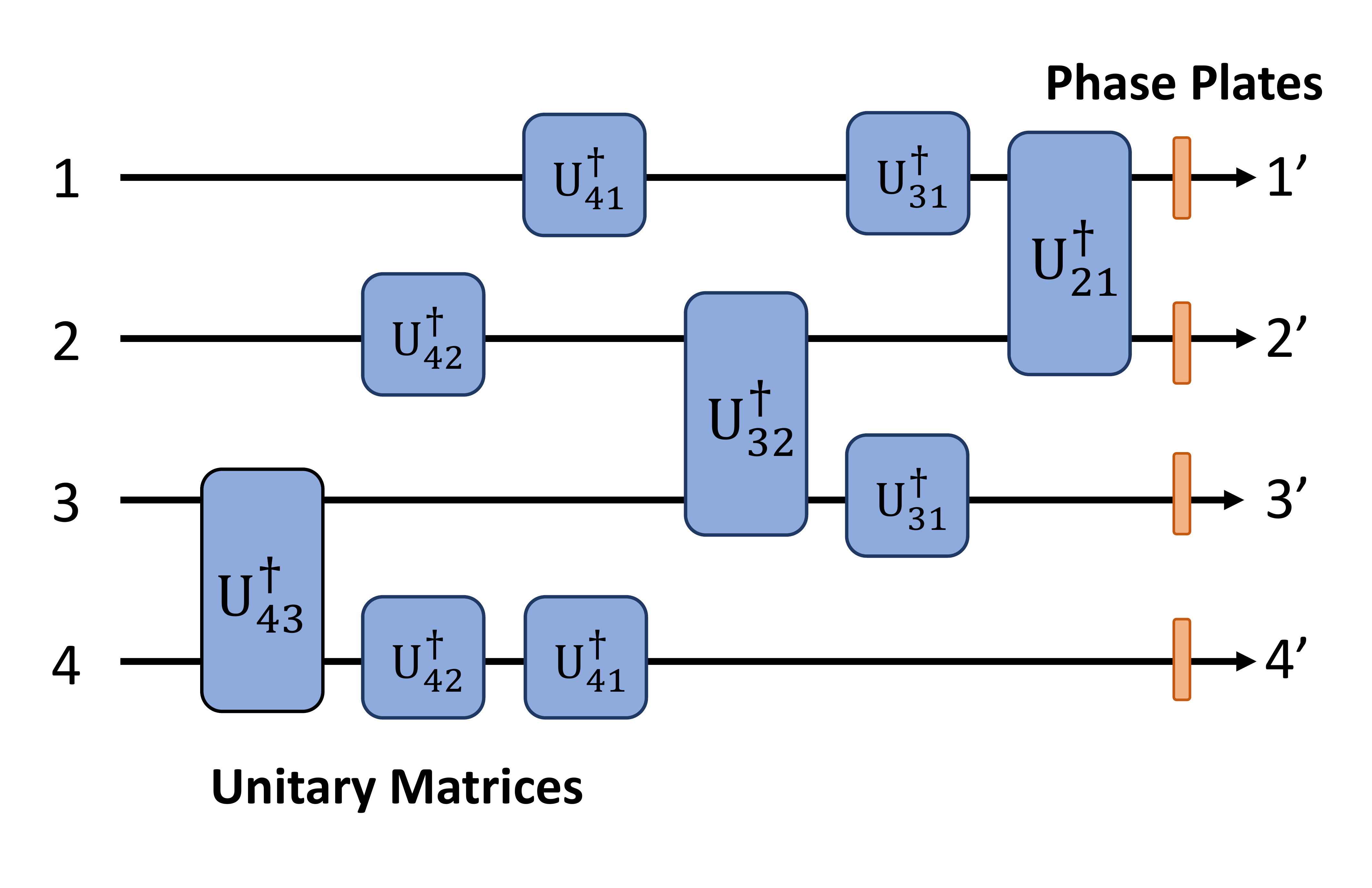}
 \caption{}\label{fig:4d_hyper}
 \end{subfigure}
 \caption{(\textbf{a}) The 4 $\times$ 4 Reck model using six beam splitters (12 with tunability). A photon can enter either one of four input ports and can leave through any of the four output ports. (\textbf{b}) The information flow representation in the case of decomposing a bulk 4 $\times$ 4 setup using 2 $\times$ 2 unitary matrices illustrated in (\textbf{a}). \label{fig:reck_four_bulk}}
\end{figure}

The matrices $T_{i,j}, i,j \in \mathbb{Z}$ are 4 $\times$ 4 matrices, which contain U(2) matrices inside. The experimental setup in the case of 3 $\times$ 3 transformation is illustrated in Figure~\ref{fig:reck_three_bulk}. The order of embedded U(2) matrices' multiplication and their action is equivalent to the physical diagram outlined. The 4 $\times$ 4 case is given in Figure~\ref{fig:reck_four_bulk}. Subfigures ({a}) and ({b}) in Figures~\ref{fig:reck_three_bulk} and~\ref{fig:reck_four_bulk} are equivalent.

\vspace{-6pt}

\subsection{Clements Decomposition Design}

\begin{figure}[H]
 \centering
 \begin{subfigure}[t]{0.5\textwidth}
 \centering
 \includegraphics[height=1in]{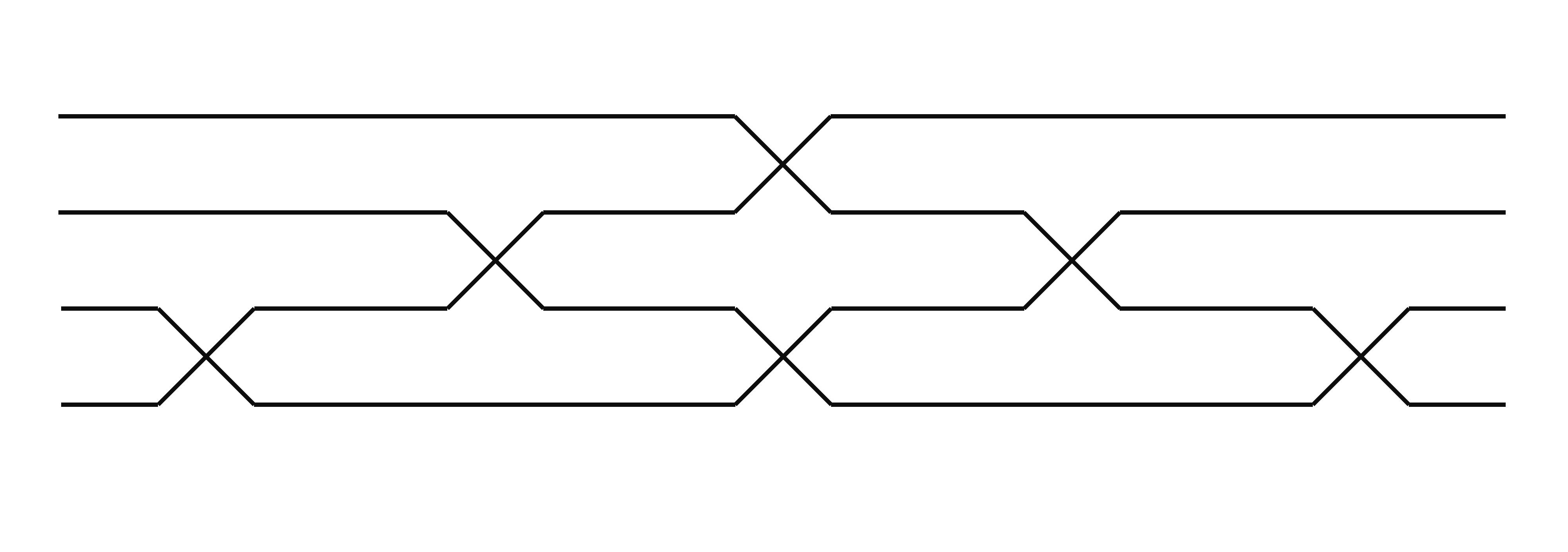}
 \caption{}
 \end{subfigure}%
 ~
 \begin{subfigure}[t]{0.5\textwidth}
 \centering
 \includegraphics[height=1in]{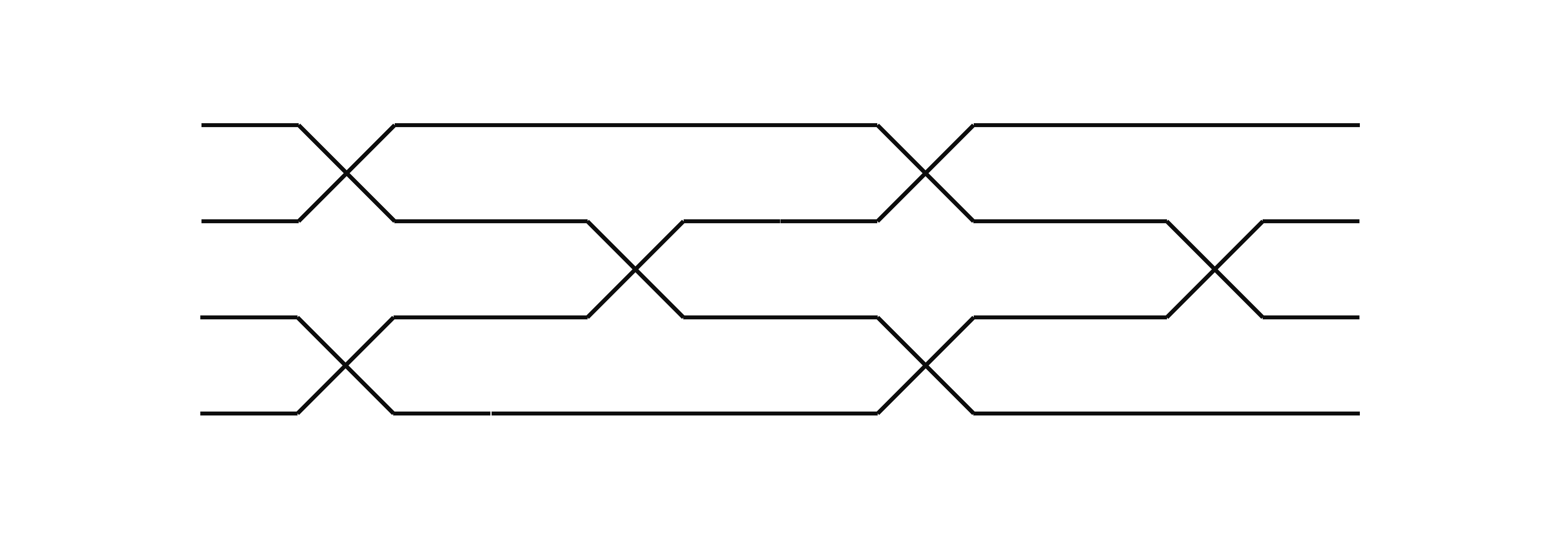}
 \caption{}
 \end{subfigure}
 \caption{(\textbf{a}) 4 $\times$ 4 mesh of the Reck design. Input photons flow from the left to the right. Each line cross-section consists of the two-dimensional two-mode mixer illustrated in Figure~\ref{fig:tunable_BS}. This is identical to the setup in Figure~\ref{fig:reck_four_bulk}a. (\textbf{b}) 4 $\times$ 4 mesh of symmetric Clements~design. \label{fig:mesh_design}}
\end{figure}

\begin{figure}[H]
\centering
\includegraphics[scale=0.2]{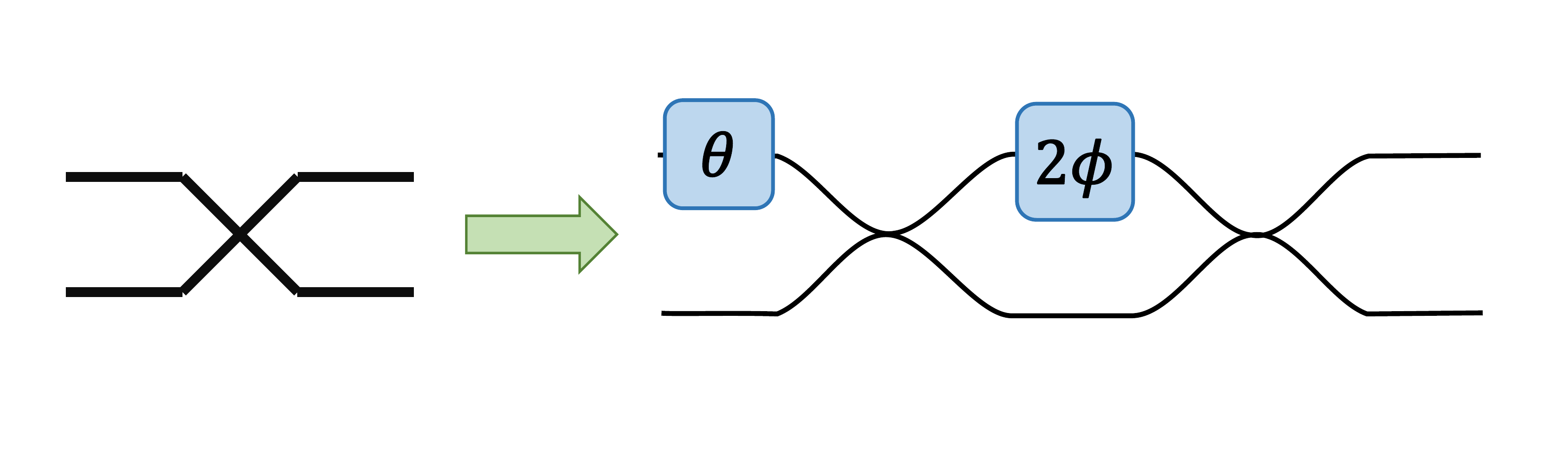}
\caption{An integrated coupler requires tunability to change a transfer matrix of the system. The tunability is acquired through an interferometer with phase shifters. $\theta$ is an external phase shift, and $\phi$ is an internal phase shift. \label{fig:tunable_BS}}
\end{figure}

The unitary matrix decomposition can also be realized in a slightly different configuration. The Clements design transforms the originally non-symmetric Reck configuration into a symmetric form~\cite{clements2016optimal}, by which we mean that the situation is non-symmetric when photons in different input ports experience different numbers of beam splitters during their propagation and before exiting the unit. It would be helpful for any future consideration to introduce a simplified mesh representation for the systems outlined in Figure~\ref{fig:mesh_design}. For example, in the case of 4 $\times$ 4 transformation, its mesh decomposition via U(2) embedded matrices could be represented either by the original decomposition proposed by Reck. The crossing parts in the mesh designs mix spatial modes and consist of integrated couplers. The tunability of the power splitting ratio can be obtained either through the dynamical change of the coupling ratio between two waveguides in an integrated coupler or by forming an interferometer using two integrated couplers executed the same task. The graphical detail of the crossing parts in mesh design is indicated in Figure~\ref{fig:tunable_BS}.

\vspace{-6pt}

The mesh designs are illustrated in Figure~\ref{fig:mesh_design}. They are equivalent to the 4 $\times$ 4 Reck model and 4~$\times$~4 Clements model. It is easy to note that a photon in the first path of the 4 $\times$ 4 Reck design illustrated in Figure~\ref{fig:mesh_design}{a} could encounter only one beam splitter, while the photon in the lowest path encounters at least three beam splitters prior to exiting the device. In the Clements symmetric design illustrated in Figure~\ref{fig:mesh_design}{b}, a photon in the first path and a photon in the last path encounter the same number of beam splitters. The loss tolerance of a quantum state becomes higher when a photon experiences the same number of beam splitter interactions~\cite{clements2016optimal}. The Clements designs are widely used instead of Reck decomposition, exactly because of its loss tolerance in quantum information~processing.

In a similar way to the Reck model, a unitary matrix can be decomposed using multiple U(2)-based matrices. Unitary matrices' realization based on the Reck model is decomposed by multiplying matrices from one side in succession. It is not necessary to multiply matrices from only one side to decompose the unitary matrix. The decomposition can be done by multiplying matrices from both sides. The U(3) case and U(4) case are given as an example.

\vspace{6pt}
\noindent U(3) case:
\begin{equation}
\begin{split}
T_{2,3}T_{1,2}U(3)T_{1,2}^{-1} = P \rightarrow U(3) = T_{1,2}^{-1}T_{2,3}^{-1}PT_{1,2}.
\end{split}
\end{equation}
U(4) case:
\begin{equation}
\begin{split}
T_{3,4}T_{2,3}U(4) T_{1,2}^{-1}T_{3,4}^{-1}T_{2,3}^{-1}T_{1,2}^{-1} = P \rightarrow U(4) = T_{2,3}^{-1}T_{3,4}^{-1}PT_{1,2}T_{2,3}T_{3,4}T_{1,2}.
\end{split}
\end{equation}

The unitary matrix decomposition is possible, and this has been experimentally realized and demonstrated~\cite{clements2016optimal,metcalf2013multiphoton}.

\subsection{Integrated Optical Tritter and Quarter}

Three- and four-dimensional directional linear optical devices will be introduced in this section. Integrated waveguide couplers~\cite{spagnolo2013three,spagnolo2013general,spagnolo2012quantum,meany2012non} can be used to implement an optical tritter illustrated in Figure~\ref{fig:tritter_integrated}{a} and its cross-section shown in Figure~\ref{fig:tritter_integrated}{b}. The propagation dynamics of such a system can be described using the same formalism as in the previous case of the directional 2 $\times$ 2 coupler:
\begin{figure}[H]
 \centering
 \begin{subfigure}[t]{0.5\textwidth}
 \centering
 \includegraphics[height=1.3in]{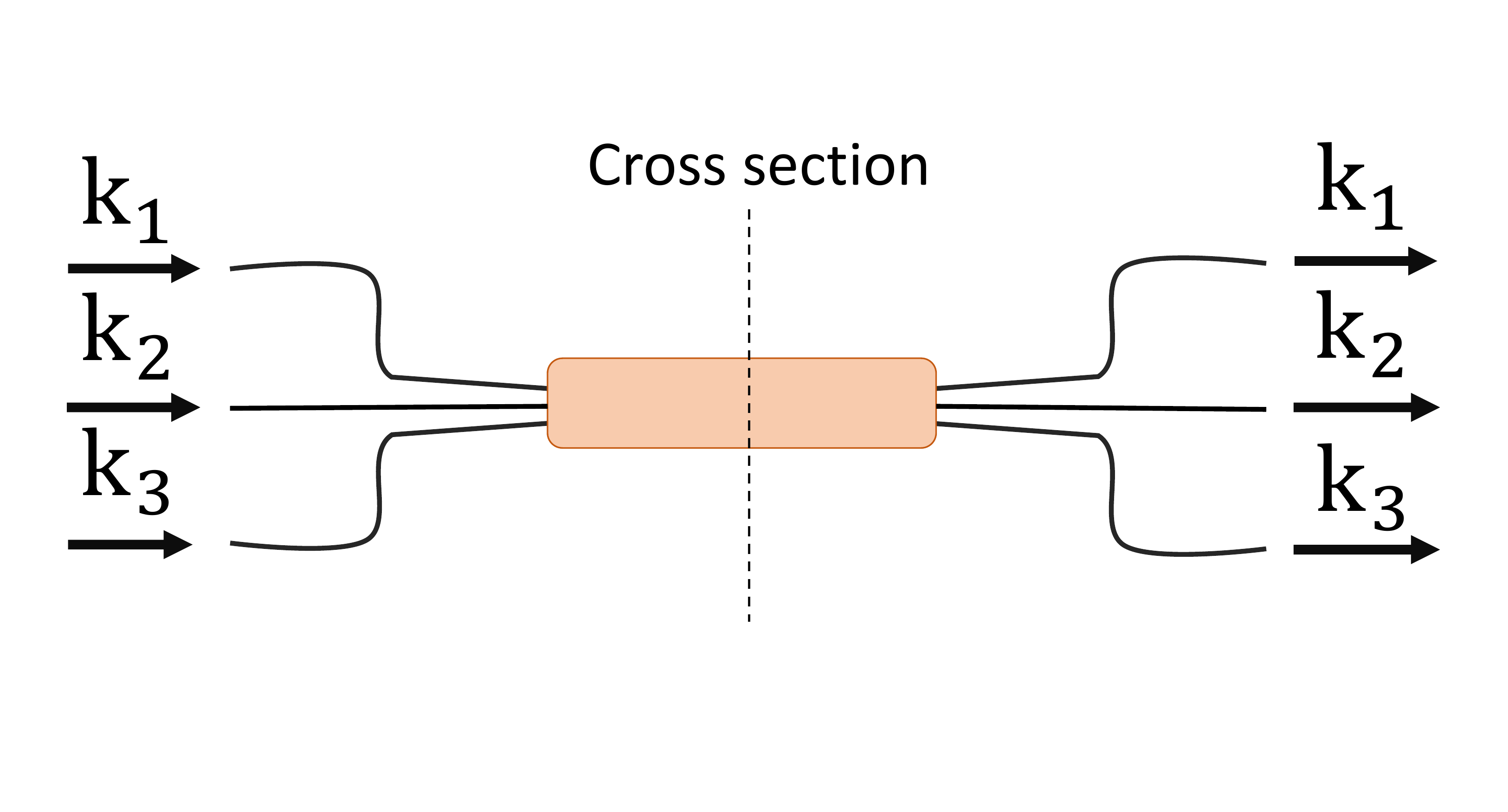}
 \caption{}\label{fig:glued_tree}
 \end{subfigure}%
 ~
 \begin{subfigure}[t]{0.5\textwidth}
 \centering
 \includegraphics[height=1.3in]{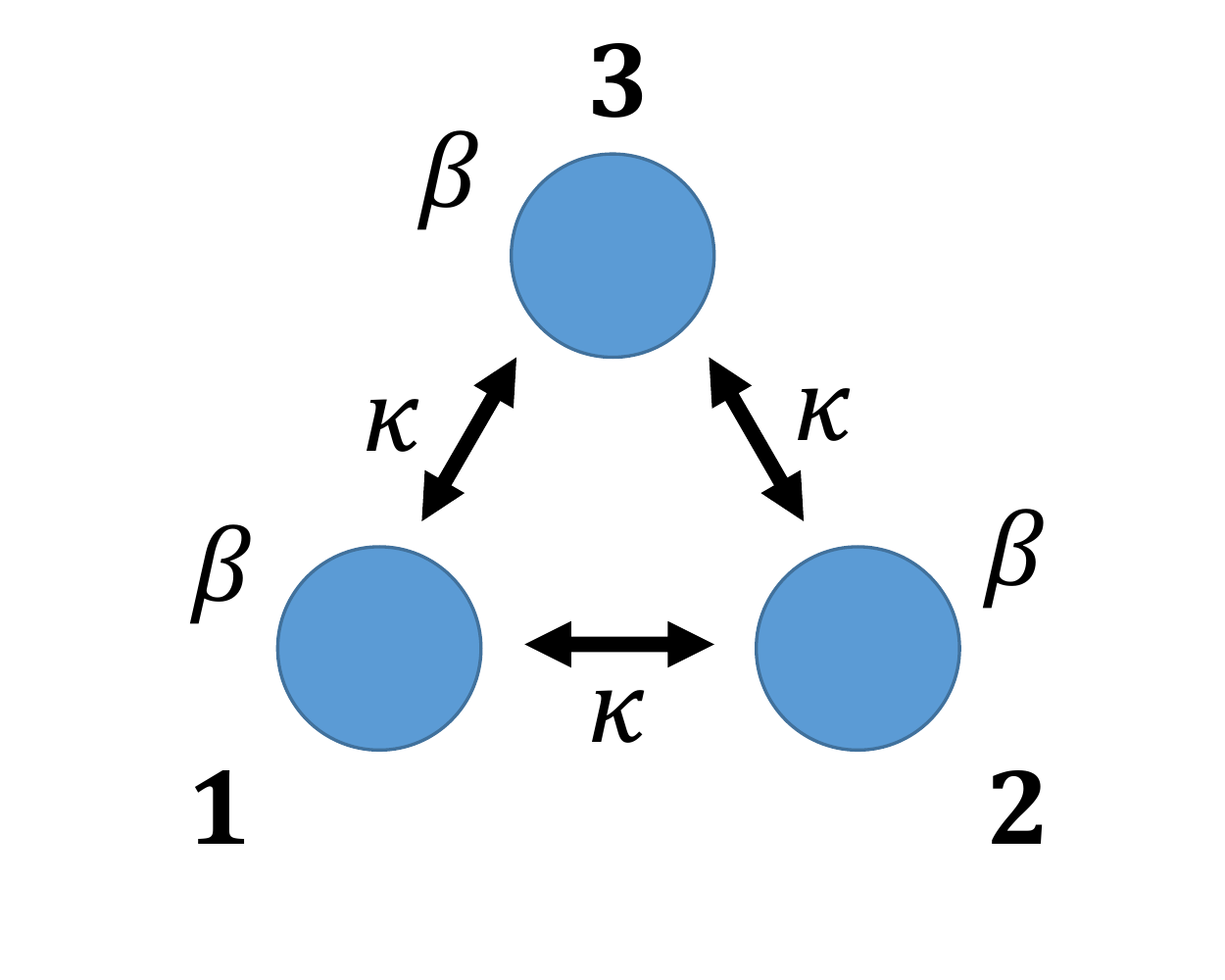}
 \caption{}\label{fig:4d_hyper}
 \end{subfigure}
 \caption{(\textbf{a}) Integrated optical tritter. $k_i$ $\{i = 1,2,3\}$ represent input and output spatial modes. (\textbf{b}) Coupling region of the integrated tritter. $\beta$ is a waveguide propagation coefficient, and $\kappa$ is a coupling coefficient between the couplers. \label{fig:tritter_integrated}}
\end{figure}

\begin{equation}
\begin{split}
 i\frac{dA_1^\dagger}{dz} = \beta A_1^\dagger + \kappa A_2^\dagger + \kappa A_3^\dagger,\\
 i\frac{dA_2^\dagger}{dz} = \kappa A_1^\dagger + \beta A_2^\dagger + \kappa A_3^\dagger,\\
 i\frac{dA_3^\dagger}{dz} = \kappa A_1^\dagger + \kappa A_2^\dagger + \beta A_3^\dagger,
\end{split}
\end{equation}

\noindent where $A_j^\dagger$ $\{j = 1,2,3\}$ are creation operators for a photon in the $j^{\text{th}}$ waveguide; $\beta$ is a propagation constant in each waveguide; and $\kappa$ is a coupling coefficient. It is assumed that all waveguides are identical and the distances between them are the same.

One could solve these equations using a matrix formalism in a similar way as the two-dimensional~case:
\begin{equation}
\begin{pmatrix}
\frac{dA_1}{dz}\\
\frac{dA_2}{dz}\\
\frac{dA_3}{dz}
\end{pmatrix}
=-i
\begin{pmatrix}
\beta & \kappa & \kappa\\
\kappa & \beta & \kappa\\
\kappa & \kappa & \beta
\end{pmatrix}
\begin{pmatrix}
A_1\\
A_2\\
A_3
\end{pmatrix}.
\end{equation}

One could find eigenvalues and corresponding eigenvectors:

\begin{equation*}
\lambda_1 = -\beta i+\kappa i:
\begin{pmatrix}
-1\\
1\\
0
\end{pmatrix}
,
\begin{pmatrix}
-1\\
0\\
1
\end{pmatrix}
,
\lambda_2 = -\beta i-2\kappa i:
\begin{pmatrix}
1\\
1\\
1
\end{pmatrix},
\end{equation*}

\begin{equation}
\begin{pmatrix}
A_1\\
A_2\\
A_3
\end{pmatrix}
=
c_1e^{-\beta z i+\kappa z i}
\begin{pmatrix}
-1\\
1\\
0
\end{pmatrix}
+c_2e^{-\beta z i+\kappa z i}
\begin{pmatrix}
-1\\
0\\
1
\end{pmatrix}
+c_3e^{-\beta z i-2\kappa z i}
\begin{pmatrix}
1\\
1\\
1
\end{pmatrix}.
\end{equation}

When the initial conditions are given by: $A_1(0) = 1, A_2(0) = 0, A_3(0) = 0$, then $c_1 = c_2 = -\frac{1}{3}, c_3 = \frac{1}{3}$,
where z is the propagation length. Other initial conditions are given as $A_1(0) = 0, A_2(0)~=~1, A_3(0) = 0$, and $A_1(0) = 0,A_2(0) = 0, A_3(0) = 1$. After solving for all initial conditions, we can obtain a total transfer matrix for the system.
\begin{equation}
U_{IntTritter} = \frac{e^{-\beta z i}}{3}
\begin{pmatrix}
2e^{\kappa z i}+e^{-2\kappa z i} & -e^{\kappa z i}+e^{-2\kappa z i} &- e^{\kappa z i}+e^{-2\kappa z i}\\
-e^{\kappa z i}+e^{-2\kappa z i}&2e^{\kappa z i}+e^{-2\kappa z i}&-e^{\kappa z i}+e^{-2\kappa z i}\\
-e^{\kappa z i}+e^{-2\kappa z i}&-e^{\kappa z i}+e^{-2\kappa z i}&2e^{\kappa z i}+e^{-2\kappa z i}
\end{pmatrix}.
\end{equation}

\begin{figure}[H]
 \centering
 \begin{subfigure}[t]{0.5\textwidth}
 \centering
 \includegraphics[height=1.3in]{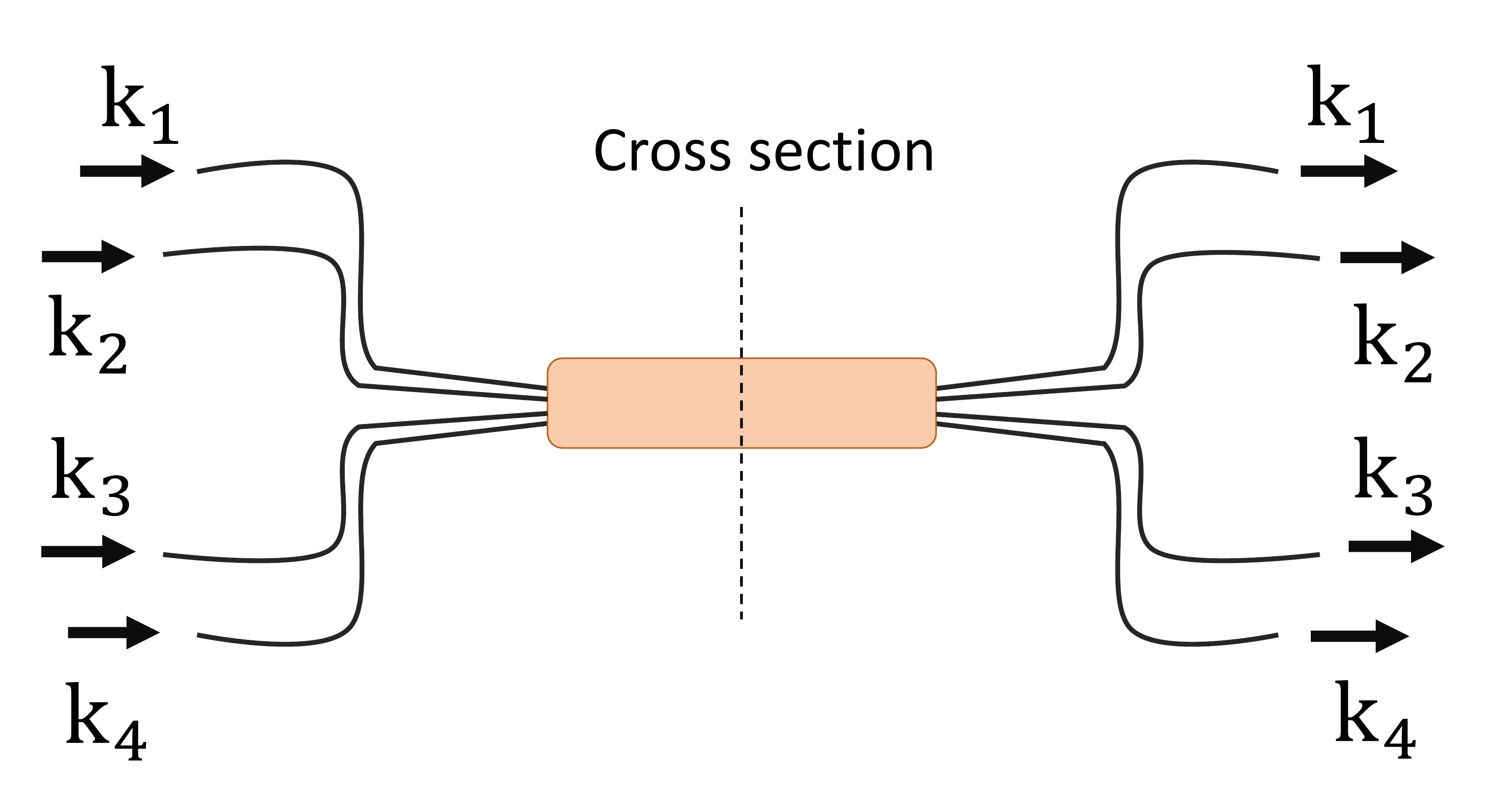}
 \caption{}
 \end{subfigure}%
 ~
 \begin{subfigure}[t]{0.5\textwidth}
 \centering
 \includegraphics[height=1.3in]{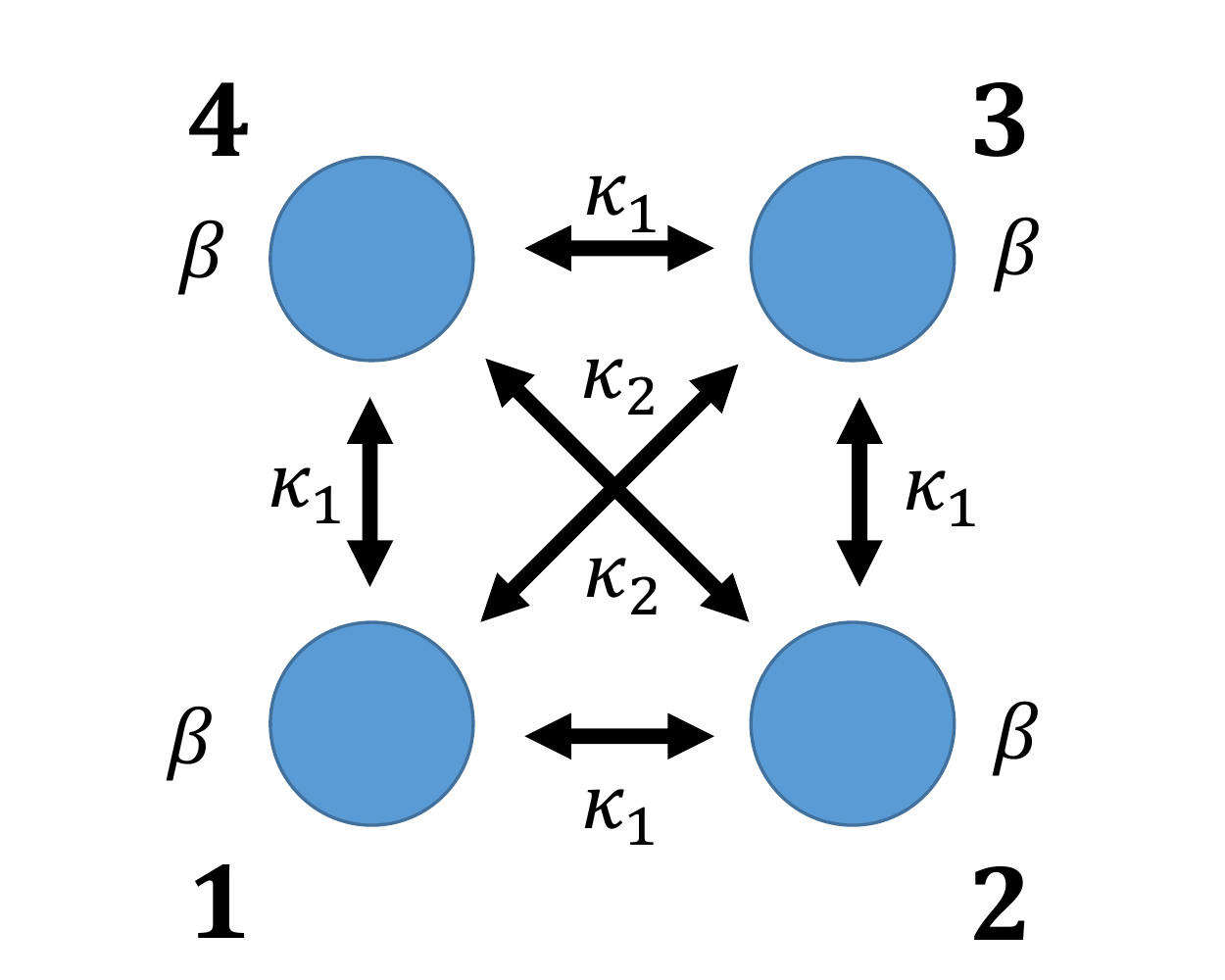}
 \caption{}
 \end{subfigure}
 \caption{(\textbf{a}) Integrated quarter. $k_i$ $\{i = 1,2,3\}$ represent input and output spatial modes. (\textbf{b}) Cross-section area of the integrated quarter. A waveguide propagation constant is $\beta$, and coupling coefficients are $\kappa_1$ for the neighboring couplers and $\kappa_2$ for the diagonal~couplers. \label{fig:quarter_integrated}}
\end{figure}

Unlike two- and three-dimensional couplers where the distances between each pair of couplers are the same, for the four-dimensional coupler with its cross-section in Figure~\ref{fig:quarter_integrated}{b}, the coupling coefficients between diagonal coupling regions are different from those on the edges of the square. We again assume the coupling strength can be controlled.
\begin{equation}
\begin{split}
 i\frac{dA_1^\dagger}{dz} = \beta A_1^\dagger + \kappa_1 A_2^\dagger + \kappa_2 A_3^\dagger+ \kappa_1 A_4^\dagger,\\
 i\frac{dA_2^\dagger}{dz} = \kappa_1 A_1^\dagger + \beta A_2^\dagger + \kappa_1 A_3^\dagger+ \kappa_2 A_4^\dagger,\\
 i\frac{dA_3^\dagger}{dz} = \kappa_2 A_1^\dagger + \kappa_1 A_2^\dagger + \beta A_3^\dagger+ \kappa_1 A_4^\dagger,\\
 i\frac{dA_4^\dagger}{dz} = \kappa_1 A_1^\dagger + \kappa_2 A_2^\dagger + \kappa_1 A_3^\dagger+ \beta A_4^\dagger,
\end{split}
\end{equation}

\noindent where $A_j^\dagger$ $\{j = 1,2,3,4\}$ are creation operators for a photon in the $j^{\text{th}}$ waveguide, $\beta$ is a propagation constant, $\kappa_1$ is a coupling coefficient between two non-diagonal couplers, and $\kappa_2$ is a coupling coefficient for two diagonal couplers.

One could solve this combination of equations using matrix formalism
\begin{equation}
\begin{pmatrix}
\frac{dA_1}{dz}\\
\frac{dA_2}{dz}\\
\frac{dA_3}{dz}\\
\frac{dA_4}{dz}
\end{pmatrix}
=-i
\begin{pmatrix}
\beta & \kappa_1 & \kappa_2 &\kappa_1\\
\kappa_1 & \beta & \kappa_1 & \kappa_2\\
\kappa_2 & \kappa_1 & \beta &\kappa_1\\
\kappa_1 & \kappa_2 & \kappa_1 &\beta
\end{pmatrix}
\begin{pmatrix}
A_1\\
A_2\\
A_3\\
A_4
\end{pmatrix}.
\end{equation}
and by finding its eigenvalues and eigenvectors:
\begin{equation}
\begin{split}
\lambda_1 = \beta i-\kappa_2 i:
\begin{pmatrix}
-1\\
0\\
1\\
0
\end{pmatrix}
,
\begin{pmatrix}
0\\
-1\\
0\\
1
\end{pmatrix},
\lambda_2 = \beta i-2\kappa_1 i+\kappa_2 i:
\begin{pmatrix}
-1\\
1\\
-1\\
1
\end{pmatrix},
\lambda_3 = \beta i+2\kappa_1 i+\kappa_2 i:
\begin{pmatrix}
1\\
1\\
1\\
1
\end{pmatrix},
\end{split}
\end{equation}
\begin{myequation1}
\begin{split}
\begin{pmatrix}
A_1\\
A_2\\
A_3\\
A_4
\end{pmatrix}
=
c_1e^{\beta z i-\kappa_2 z i}
\begin{pmatrix}
-1\\
0\\
1\\
0
\end{pmatrix}
+c_2e^{\beta z i-\kappa_2 z i}
\begin{pmatrix}
0\\
-1\\
0\\
1
\end{pmatrix}
+c_3e^{\beta z i-2\kappa_1 z i+\kappa_2 z i}
\begin{pmatrix}
-1\\
1\\
-1\\
1
\end{pmatrix}
+c_4e^{\beta z i+2\kappa_1 z i+\kappa_2 z i}
\begin{pmatrix}
1\\
1\\
1\\
1
\end{pmatrix}.
\end{split}
\end{myequation1}

When the initial conditions are given by: $A_1(0) = 1,A_2(0) = 0, A_3(0) = 0,A_4(0) = 0$, then $c_1 = -\frac{1}{2}, c_2 = 0, c_3 = -\frac{1}{4}, c_4 = \frac{1}{4}$
where z is the propagation length. 
After imposing the initial conditions and solving, we obtain a total transfer matrix for the system:
\begin{equation}
U_{IntQuarter} = \frac{e^{\beta z i}}{4}
\begin{pmatrix}
A & B & C & B\\
B & C & B & A\\
C & B & A & B\\
B & A & B & C
\end{pmatrix},
\end{equation}
where:
\begin{equation}
\begin{split}
A& = 2e^{-\kappa_2 z i}+e^{(-2\kappa_1 +\kappa_2)z i}+e^{(2\kappa_1 +\kappa_2)z i},\\
B &= -e^{(-2\kappa_1 +\kappa_2)z i}+e^{(2\kappa_1 +\kappa_2)z i}, \\
C &= -2e^{-\kappa_2 z i}+e^{(-2\kappa_1 +\kappa_2)z i}+e^{(2\kappa_1 +\kappa_2)z i}. \\
\end{split}
\end{equation}

\section{Directionally-Unbiased Linear-Optical Designs}
\textls[-20]{Directional devices were introduced in the previous sections. In this section, {directionally-unbiased}} optical devices will be introduced. Some types of directionally-unbiased systems have existed in the literature for a long time~\cite{zhang2000investigation,saleh1988reflective}, while new beam splitter-based designs have been introduced recently~\cite{simon2016group,osawa2018experimental}. The new beam splitter-based design will be reviewed first, and then, a directionally-unbiased design based on an optical tritter will be introduced. These devices are the building blocks for quantum information processing applications, especially quantum walks. Three- and four-port designs will be reviewed in the following subsections.

\subsection{Directionally-Unbiased Linear-Optical Three- and Four-Port Devices} \label{multiports_sub}

\begin{figure}[H]
\centering
\includegraphics[scale=0.2]{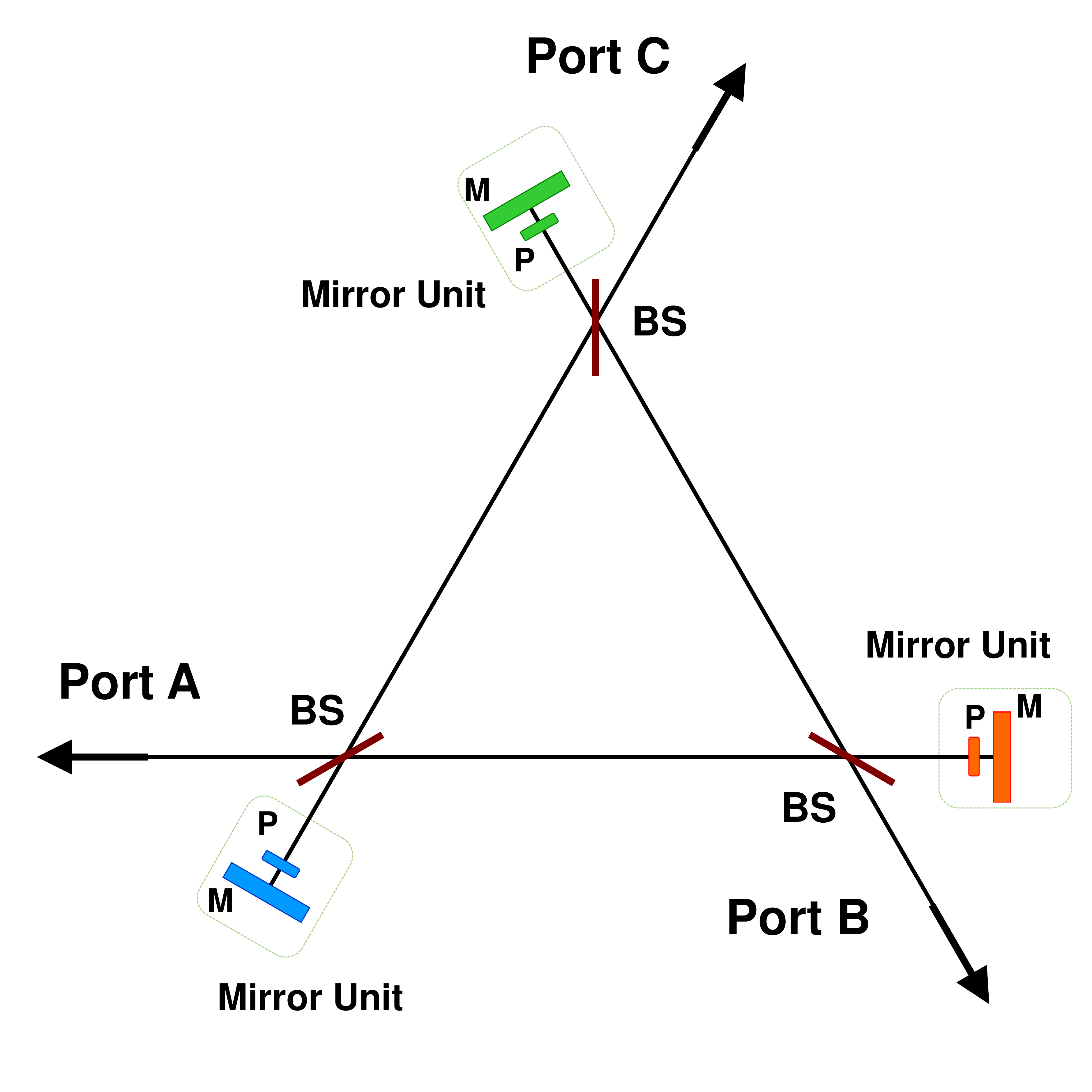}
\caption{Directionally-unbiased linear-optical three-port operation. Input ports can be used as outputs in this system. For example, if a photon is inserted at Port A, then the photon would leave Ports A, B, and C. The beam splitter (BS) splits the incoming photon into two outgoing directions. Mirror units are necessary to reverse the propagation direction so that the directional system becomes directionally-unbiased. Mirror units consist of a phase shifter P and mirror M, as illustrated in the~figure. \label{fig:3x3_multiport}}
\end{figure}

\begin{figure}[H]
\centering
\includegraphics[scale=0.2]{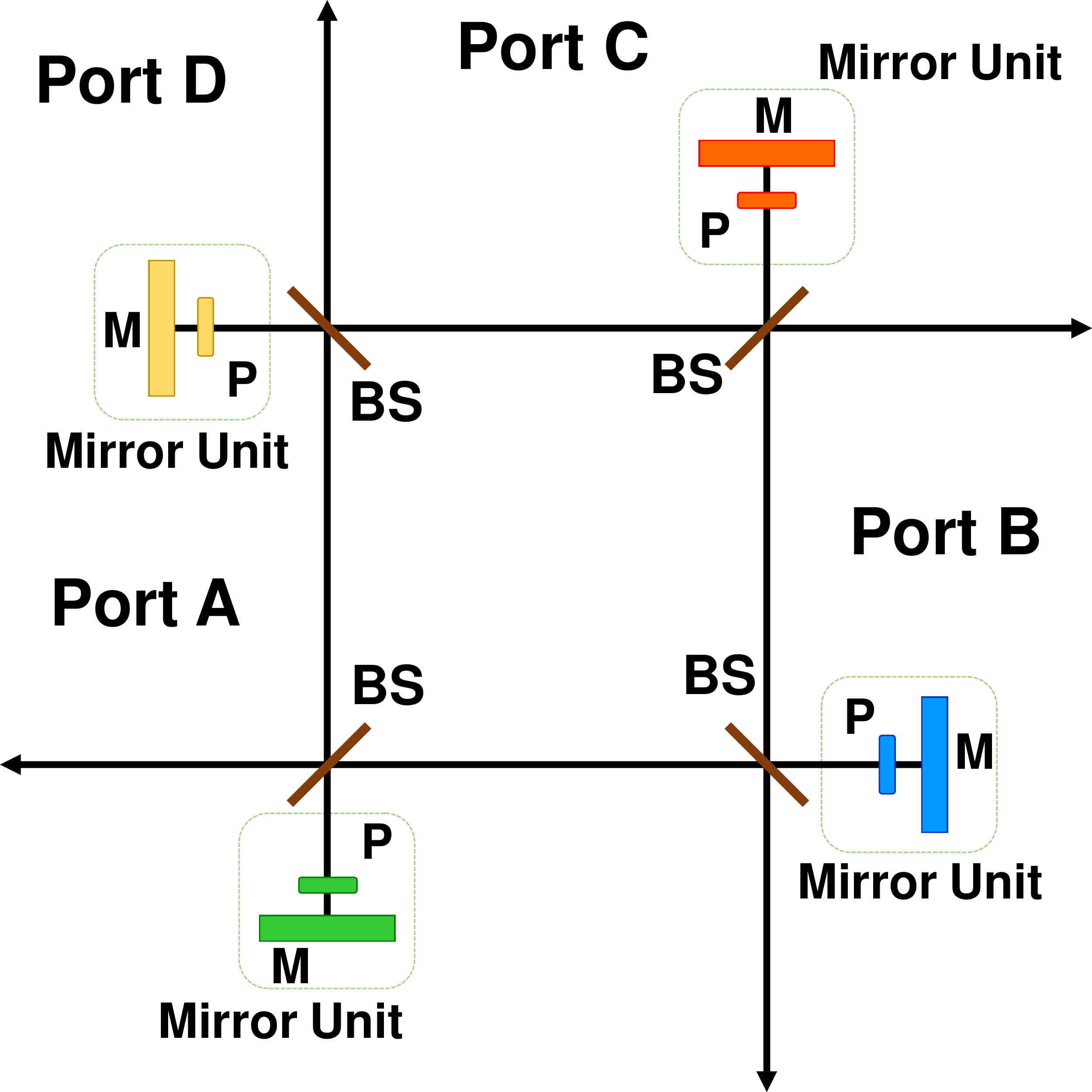}
\caption{Directionally-unbiased linear-optical four-port operation. It has the same configuration as Figure~\ref{fig:3x3_multiport} with one extra input/output port.\label{fig:4x4_multiport}}
\end{figure}

Recently, several new designs for directionally-unbiased linear-optical multiports have been introduced, and they have applications in quantum simulations of Hamiltonians and topological phase simulations~\cite{simon2017quantum2,simon2017quantum1,simon2018joint,simon2016group}. A three-port operation has been experimentally demonstrated using bulk optical devices~\cite{osawa2018experimental}. The basic components of the multiport devices are beam splitters, mirrors, and phase shifters. The three-port operation is given in Figure~\ref{fig:3x3_multiport}. The entrance and exit ports are denoted as Port A, Port B, and Port C. A photon can enter any of the three ports, and the photon can leave any of the three ports. When a photon enters the system, the photon amplitude will be split at each beam splitter in the system. The final output photon amplitudes can be described by a transfer matrix constructed by adding all the possible paths the photon would take before it leaves the system. The coherence of the photon needs to be long enough to add all amplitudes coherently within the system. The four ports illustrated in Figure~\ref{fig:4x4_multiport} will be introduced, as well as three ports in this subsection. 

\vspace{-6pt}

All input-output transfer elements for three ports could be considered as a coherent superposition of all possible paths inside the device:
\begin{equation}
\begin{aligned}
A \rightarrow A = {} & \frac{1}{4}e^{i\phi_C}+\frac{1}{4}e^{i\phi_B}-i\frac{1}{8}e^{i(\phi_B+\phi_C)}-i\frac{1}{8}e^{i(\phi_B+\phi_C)}+\frac{1}{16}e^{i(\phi_A+\phi_B+\phi_C)}+\frac{1}{16}e^{i(\phi_A+\phi_B+\phi_C)}\\
&-\frac{1}{16}e^{i(\phi_B+\phi_C+\phi_B)}-\frac{1}{16}e^{i(\phi_C+\phi_B+\phi_C)}-\frac{1}{16}e^{i(\phi_B+\phi_A+\phi_B)}-\frac{1}{16}e^{i(\phi_C+\phi_A+\phi_C)} + \dots,
\end{aligned}\label{eqn:transitionA}
\end{equation}
\begin{equation}
\begin{aligned}
A \rightarrow B = {} & i\frac{1}{2}-\frac{1}{4}e^{i\phi_C}-i\frac{1}{8}e^{i(\phi_A+\phi_B)}
+i\frac{1}{8}e^{i(\phi_A+\phi_C)}+i\frac{1}{8}e^{i(\phi_B+\phi_C)}+\frac{1}{16}e^{i(\phi_A+\phi_B+\phi_C)}\\
&-\frac{1}{16}e^{i(\phi_A+\phi_B+\phi_C)}-\frac{1}{16}e^{i(\phi_A+\phi_B+\phi_C)}+\frac{1}{16}e^{i(\phi_C+\phi_B+\phi_C)}+\frac{1}{16}e^{i(\phi_C+\phi_A+\phi_C)} +\dots,
\end{aligned}
\end{equation}\label{eqn:transitionB}
\begin{equation}
\begin{aligned}
A \rightarrow C = {} & i\frac{1}{2}-\frac{1}{4}e^{i\phi_B}+i\frac{1}{8}e^{i(\phi_A+\phi_B)}-i\frac{1}{8}e^{i(\phi_A+\phi_C)}+i\frac{1}{8}e^{i(\phi_B+\phi_C)}+\frac{1}{16}e^{i(\phi_A+\phi_B+\phi_C)}\\
&-\frac{1}{16}e^{i(\phi_A+\phi_B+\phi_C)}-\frac{1}{16}e^{i(\phi_A+\phi_B+\phi_C)}+\frac{1}{16}e^{i(\phi_B+\phi_C +\phi_B)}+\frac{1}{16}e^{i(\phi_B+\phi_A+\phi_B)} +\dots .
\end{aligned}
\end{equation}\label{eqn:transitionC}

$A \rightarrow A$ represents a transfer amplitude for the input A back to the output A. A similar geometric sum can be constructed for elements describing the photon coming in at B and leaving through Ports A, B, and C and, similarly, for an input C and the output through A, B, and C. Using the values above, a transfer matrix for this system can be reconstructed:

\begin{equation}
U_{multiport}
=
\begin{pmatrix}
U_{A \rightarrow A} & U_{B \rightarrow A}& U_{C \rightarrow A}\\
U_{A \rightarrow B}& U_{B \rightarrow B}& U_{C \rightarrow B}\\
U_{A \rightarrow C}& U_{B \rightarrow C}& U_{C \rightarrow C}
\end{pmatrix}.
\end{equation}

Here, $U_{A\rightarrow A}$ represents a transition from Port A to Port A. All terms are coherently summed; therefore, these transition amplitudes are for the long-time limit. The rest of the elements describe other possible (input $\rightarrow$ output) transitions.

The dimensionality and the number of optical elements is increased in the case of a four-port device as illustrated in Figure~\ref{fig:4x4_multiport}. It needs to be noted that four ports are slightly different from the three ports because of the numbers of beam splitter encounters before the input photon leaves the system. For example, the shortest path for Port A to Port B would be $A \rightarrow B$ with one beam splitter encounter. Similarly, the shortest path for Port A to Port C would be $A \rightarrow B \rightarrow C$ or $A \rightarrow D \rightarrow C$ with two beam splitter encounters. The probability amplitude is lower for the path A to C because the photon encounters one extra beam splitter. This path-dependent amplitude difference needs to be considered for higher dimensional multiport implementation.
\begin{equation}
\begin{aligned}
A \rightarrow A &= \frac{1}{4}e^{i\phi_B}+\frac{1}{4}e^{i\phi_D}-\frac{1}{16}e^{i(\phi_B+\phi_C+\phi_B)}+\frac{1}{16}e^{i(\phi_B+\phi_C+\phi_D)}-\frac{1}{16}e^{i(\phi_D+\phi_C+\phi_D)}\\
&+\frac{1}{16}e^{i(\phi_D+\phi_C+\phi_B)}-\frac{1}{16}e^{i(\phi_B+\phi_A+\phi_B)}-\frac{1}{16}e^{i(\phi_D+\phi_A+\phi_D)} + \dots,
\end{aligned}\label{eqn:transitionA4}
\end{equation}

\begin{equation}
\begin{aligned}
A \rightarrow B &= i\frac{1}{2}-i\frac{1}{8}e^{i(\phi_D+\phi_C)}-i\frac{1}{8}e^{i(\phi_B+\phi_A)}+i\frac{1}{8}e^{i(\phi_B+\phi_C)}+i\frac{1}{8}e^{i(\phi_D+\phi_A)}+ \dots,
\end{aligned}\label{eqn:transitionB4}
\end{equation}

\begin{equation}
\begin{aligned}
A \rightarrow C &= -\frac{1}{4}e^{i\phi_B}-\frac{1}{4}e^{i\phi_D}-\frac{1}{16}e^{i(\phi_B+\phi_C+\phi_D)}+\frac{1}{16}e^{i(\phi_B+\phi_A+\phi_B)}+\frac{1}{16}e^{i(\phi_D+\phi_C+\phi_D)}\\
&-\frac{1}{16}e^{i(\phi_D+\phi_C+\phi_B)}+\frac{1}{16}e^{i(\phi_B+\phi_A+\phi_B)}+\frac{1}{16}e^{i(\phi_D+\phi_A+\phi_D)} + \dots,
\end{aligned}\label{eqn:transitionC4}
\end{equation}

\begin{equation}
\begin{aligned}
A \rightarrow D &= i\frac{1}{2}-i\frac{1}{8}e^{i(\phi_B+\phi_C)}-i\frac{1}{8}e^{i(\phi_D+\phi_C)}-i\frac{1}{8}e^{i(\phi_D+\phi_A)}+i\frac{1}{8}e^{i(\phi_B+\phi_A)}+ \dots.
\end{aligned}\label{eqn:transitionD4}
\end{equation}

The final transfer matrix consists of 16 (input $\rightarrow$ output) transition amplitudes:

\begin{equation}
U_{multiport}
=
\begin{pmatrix}
U_{A \rightarrow A} & U_{B \rightarrow A}& U_{C \rightarrow A}&U_{D \rightarrow A}\\
U_{A \rightarrow B}& U_{B \rightarrow B}& U_{C \rightarrow B}&U_{D \rightarrow B}\\
U_{A \rightarrow C}& U_{B \rightarrow C}& U_{C \rightarrow C}&U_{D \rightarrow C}\\
U_{A \rightarrow D}& U_{B \rightarrow D}& U_{C \rightarrow D}&U_{D \rightarrow D}
\end{pmatrix}.
\end{equation}

\subsection{Constructing Reversible Optical Tritter and Quarter}

\begin{figure}[H]
 \centering
 \begin{subfigure}[t]{0.5\textwidth}
 \centering
 \includegraphics[height=1in]{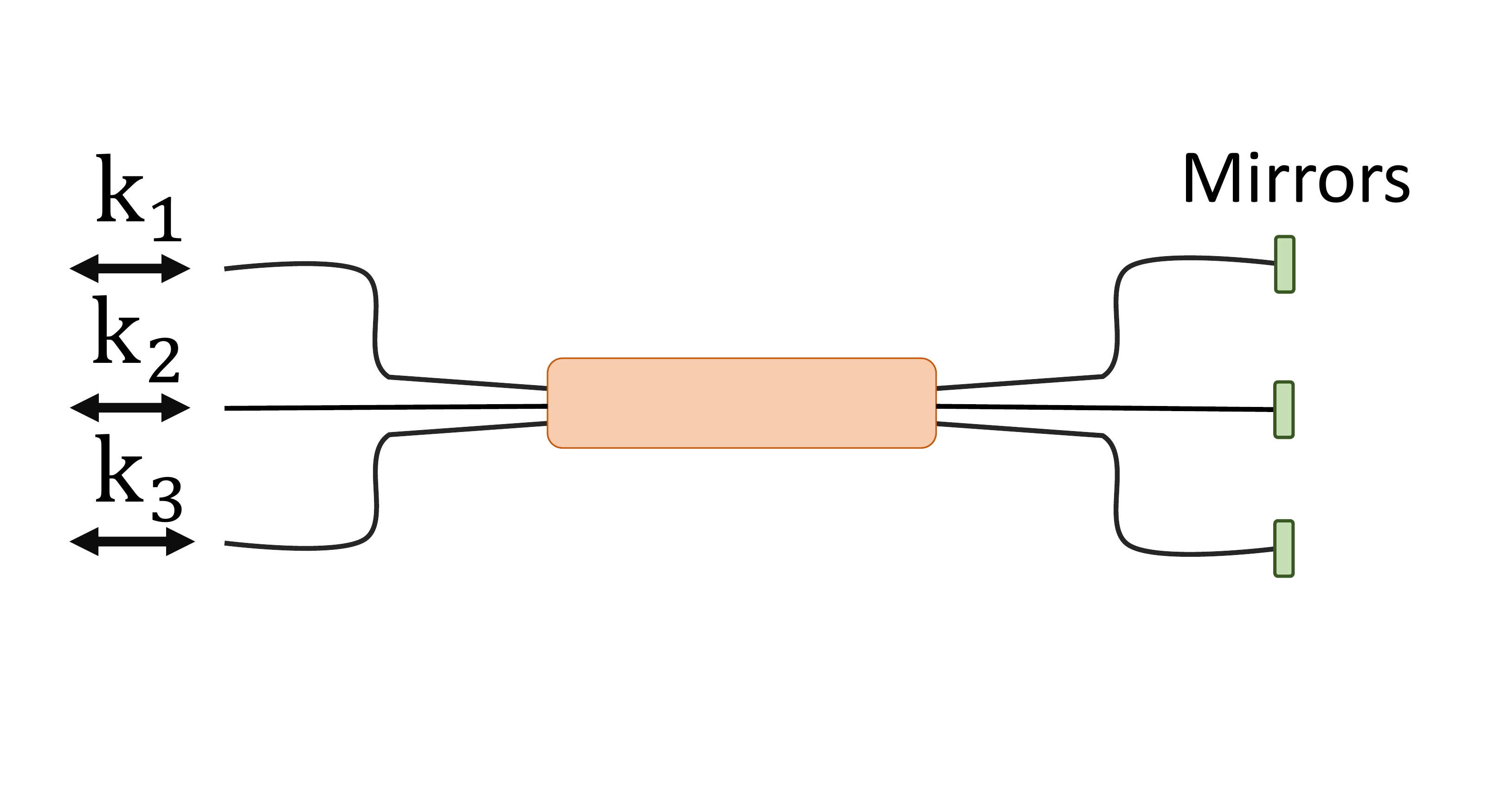}
 \caption{}\label{fig:glued_tree}
 \end{subfigure}%
 ~
 \begin{subfigure}[t]{0.5\textwidth}
 \centering
 \includegraphics[height=1in]{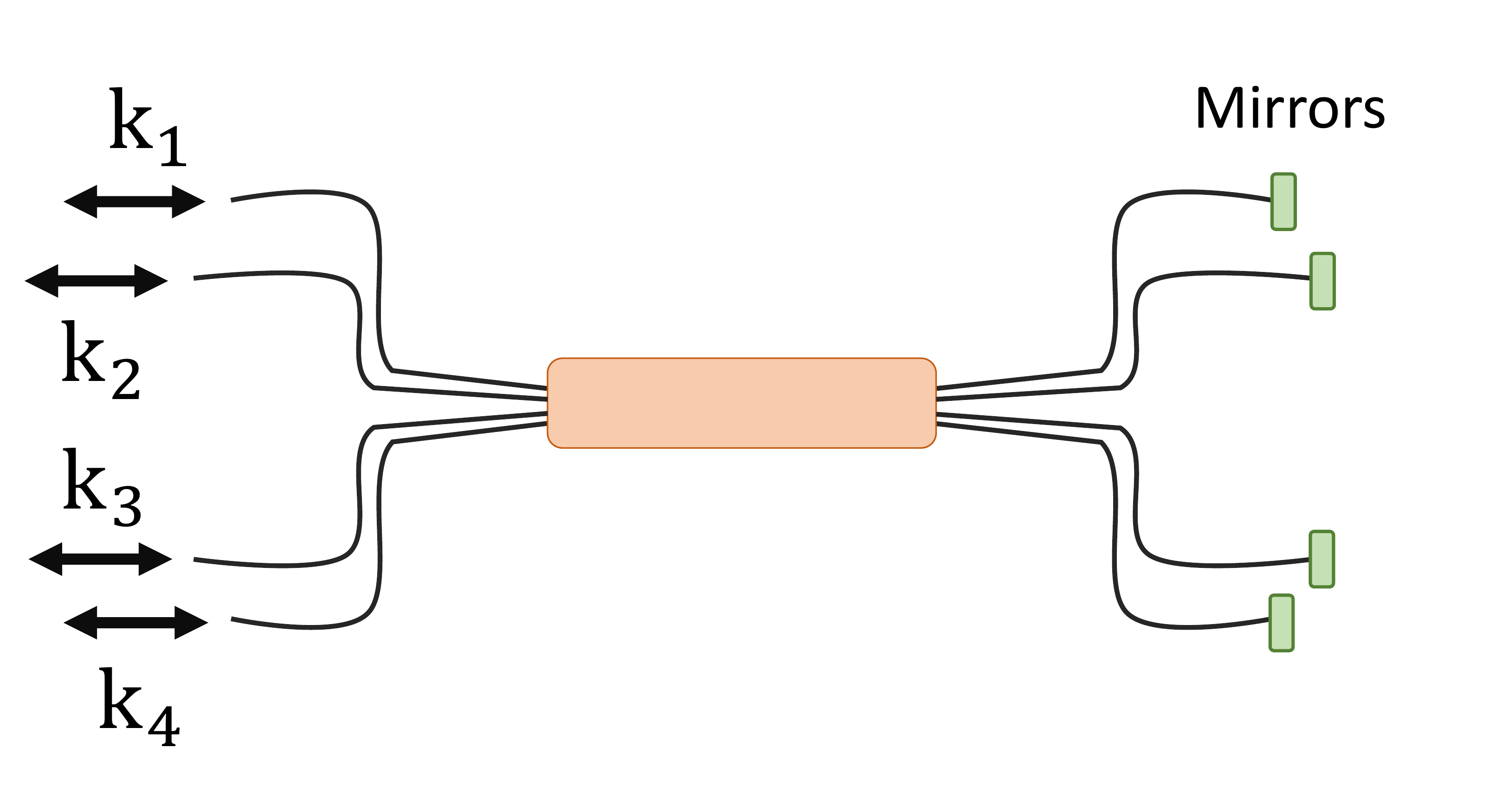}
 \caption{}\label{fig:4d_hyper}
 \end{subfigure}
 \caption{(\textbf{a}) Reversible integrated tritter. The photon propagation direction is reversed by placing mirrors at the end of the coupling region. (\textbf{b}) Reversible integrated quarter. It has the same configuration as (\textbf{a}) with an extra input and output port. \label{fig:rev_tritter}}
\end{figure}

It has been shown earlier that traditional integrated optical 3 $\times$ 3 and 4 $\times$ 4 couplers are directionally-biased devices. However, placing mirrors at each of the output ports of the device helps to eliminate this directional bias and return the optical signal back to any of the input ports. This type of reversible design has been also introduced in the area of linear interferometric networks and has been sometimes called a generalized Michelson interferometer~\cite{vance1995general,schwelb1998generalized}. The reversibility is introduced by mirrors, while additional phase shifters can be introduced before the mirrors. Both 3 $\times$ 3 and 4 $\times$ 4 implementations are illustrated in Figure~\ref{fig:rev_tritter}. The reversed tritter can be realized using the same design formalism as a generalized Michelson interferometer. The input photon state is transformed by a tritter matrix, phase shifters, and a transposed tritter matrix.

\begin{equation}
\begin{split}
U_{RevTritter} =
U_{IntTritter}^T
U_{phase}
U_{IntTritter},
\end{split}
\end{equation}

\begin{equation}
U_{phase}=
\begin{pmatrix}
e^{i\phi_A} & 0 & 0\\
0 & e^{i\phi_B} & 0\\
0 & 0 & e^{i\phi_C}
\end{pmatrix},
\end{equation}
\begin{equation}
U_{IntTritter} = \frac{e^{-\beta z i}}{3}
\begin{pmatrix}
2e^{\kappa z i}+e^{-2\kappa z i} & -e^{\kappa z i}+e^{-2\kappa z i} &- e^{\kappa z i}+e^{-2\kappa z i}\\
-e^{\kappa z i}+e^{-2\kappa z i}&2e^{\kappa z i}+e^{-2\kappa z i}&-e^{\kappa z i}+e^{-2\kappa z i}\\
-e^{\kappa z i}+e^{-2\kappa z i}&-e^{\kappa z i}+e^{-2\kappa z i}&2e^{\kappa z i}+e^{-2\kappa z i}
\end{pmatrix}
=
\frac{e^{-\beta z i}}{3}
\begin{pmatrix}
A & B & B\\
B & A & B\\
B & B & A
\end{pmatrix},
\end{equation}
\noindent where $A = 2e^{\kappa z i}+e^{-2\kappa z i}$ and $B=-e^{\kappa z i}+e^{-2\kappa z i}.$ Putting all the pieces together, the final expression is:
\begingroup\makeatletter\def\f@size{9}\check@mathfonts
\def\maketag@@@#1{\hbox{\m@th\fontsize{10}{10}\selectfont\normalfont#1}}%
\begin{myequation2}
U_{RevTritter}=
\frac{e^{-2\beta z i}}{9}
\begin{pmatrix}
e^{i\phi_A}AA + e^{i\phi_B}BB + e^{i\phi_C}BB & e^{i\phi_A}AB + e^{i\phi_B}BA + e^{i\phi_C}BB & e^{i\phi_A}AB + e^{i\phi_B}BB + e^{i\phi_C}BA\\
e^{i\phi_A}BA + e^{i\phi_B}AB + e^{i\phi_C}BB & e^{i\phi_A}BB + e^{i\phi_B}AA + e^{i\phi_C}BB & e^{i\phi_A}BB + e^{i\phi_B}AB + e^{i\phi_C}BA\\
e^{i\phi_A}BA + e^{i\phi_B}BB + e^{i\phi_C}AB & e^{i\phi_A}BB + e^{i\phi_B}BA + e^{i\phi_C}AB & e^{i\phi_A}BB + e^{i\phi_B}BB + e^{i\phi_C}AA
\end{pmatrix}.
\end{myequation2}
\endgroup
\begin{equation}
U_{phase}=
\begin{pmatrix}
e^{i\phi_A} & 0 & 0 & 0\\
0 & e^{i\phi_B} & 0 & 0\\
0 & 0 & e^{i\phi_C} & 0\\
0 & 0 & 0 & e^{i\phi_D}
\end{pmatrix},
\end{equation}
\noindent where $\phi_{A},\phi_{B},\phi_{C}$ and $\phi_{D}$ are phase shifts introduced before the second device encounter.
\begin{equation}
U_{IntQuarter}=\frac{e^{\beta z i}}{4}
 \begin{pmatrix}
 A & B & C & B\\
 B & C & B & A\\
 C & B & A & B\\
 B & A & B & C
 \end{pmatrix},
\end{equation}
\noindent where $A = 2e^{-\kappa_2 z i}+e^{-2\kappa_1 z i+\kappa_2 z i}+e^{2\kappa_1 z i+\kappa_2 z i}$, $B=-e^{-2\kappa_1 z i+\kappa_2 z i}+e^{2\kappa_1 z i+\kappa_2 z i}$, and $C = -2e^{-\kappa_2 z i}+e^{-2\kappa_1 z i+\kappa_2 z i}+e^{2\kappa_1 z i+\kappa_2 z i}.$
The reversible quarter matrix is then derived from the equation below.
\begin{equation}
\begin{split}
U_{RevQuarter} =
U_{IntQuarter}^T
U_{phase}
U_{IntQuarter}.
\end{split}
\end{equation}

\section{Discrete-Time Quantum Walks} \label{quantum_walks}
Up to this section, photonics-based linear optical devices have been investigated. A major motivation for focusing on the directionally-unbiased versions is in their potential for implementing quantum walk applications. Quantum walks are motivated from classical random walks. Quantum walks can support superposition states and interference in the system where interference is absent in classical random walks. There are two types of quantum walks, discrete-time quantum walks and continuous-time quantum walks. During the discrete-time quantum walk, the evolution operator is applied in a discrete time fashion while the operator application timing is irrelevant in the continuous case. We focus on the discrete case in this review. The simplest classical and quantum walk design would be a walk performed on a line. In the case of a classical random walk on a line with an unbiased two-dimensional coin, a walker can hop one step to the right or to the left with equal probability depending on the result of a coin-toss event. The walker walks on a line for certain steps, and the probability at a specific position can be obtained by repeating the process. By recording all the probability at each location on the line, a probability distribution associated with that coin is constructed. Classical random walks involve intermediate measurement, meaning the position of the walker is measured right after a coin-toss event. In contrast, the quantum approach to random walks preserves the coherence of all possible paths by not measuring an intermediate state of the walker and, as a consequence, enables the quantum interference of available probability amplitudes. It is known that the probability distribution spreads faster in quantum walks compared to classical random walks.
Classical random walks are useful for many randomized algorithm implementations~\cite{motwani1996randomized}. It is natural to consider that quantum walks could achieve better outcomes than classical random walk-based algorithms, and it is indeed possible to gain algorithmic speedup using the fact that quantum walks can spread faster than classical random walks. Several different algorithms have been developed through quantum walks, and some are faster than classical algorithms. Hitting time, graph traversal speed from a point to another point in a graph, on a hypercube~\cite{moore2002quantum,krovi2006hitting}, and a glued tree are known to be exponentially faster in the quantum case~\cite{childs2002example,childs2003exponential}. Element distinctness~\cite{ambainis2007quantum}, triangle finding~\cite{magniez2007quantumtri}, matrix product verification~\cite{buhrman2006quantum}, and group commutativity testing~\cite{magniez2007quantumcom} have been also investigated. Flexible graph construction is necessary to perform quantum walk-based algorithms. Any graphs consist of vertices and edges, and these need to be prepared in an experimentally realizable way. This task can be achieved through linear-optical devices, which have several input and output ports as discussed in previous sections. An experimental quantum walk implementation has been demonstrated in optical systems using optical cavities~\cite{bouwmeester1999optical,knight2003optical}, optical rings~\cite{knight2003quantum}, time-bins~\cite{schreiber2010photons}, Michelson interferometers~\cite{pandey2011quantum}, optical network~\cite{zhao2002implement}, beam displacers~\cite{broome2010discrete}, orbital angular momentum manipulation~\cite{goyal2013implementing,zhang2007demonstration,cardano2015quantum}, and optical refraction~\cite{francisco2006simulating}. The majority of these implementations are based on directional-optical devices; therefore, their implementation costs would rapidly increase as the dimensionality of the quantum walk system becomes higher. This applies to spatially-multiplexed quantum walk systems as they need to use beam splitters in a feed-forward manner. Time-multiplexed quantum walks are also commonly used since they can be compact. However, it would be challenging to perform node-by-node amplitude tuning. Integrated waveguide-based systems can be made directionally-unbiased and have been experimentally demonstrated~\cite{peruzzo2010quantum,tang2018experimental,tang2018experimental2}. Directionally-unbiased linear-optical multiport-based quantum walk configurations, which can realize amplitude tunability while offering an implementation resource reduction, will be introduced in the following several subsections.

\subsection{Coin Walk: Quantum Walk on Vertices}

The traditional quantum walk is illustrated using a position Hilbert space $H_P$ and a ``coin'' Hilbert space $H_C$. A quantum walker's position is described by the amplitudes in a position space spanned by $\{\ket{m}, m \in \mathbb{Z}\}$, and a coin space is spanned by a two-dimensional computational basis $\{\ket{R} \equiv (1,0)^T,\ket{L} \equiv (0,1)^T\}.$ The Hilbert space of the system is given by $H = H_P \otimes H_C$. We define a coin operator $\hat{C}$ and a shift operator $\hat{S}$ acting on each Hilbert space. The shift operator translates a walker's position from $\ket{m}$ to $\ket{m-1}$ or $\ket{m+1}$ depending on the result of the coin operation.
\begin{equation}
\hat{S}\ket{m}\ket{R} = \ket{m+1}\ket{R} \ \mbox{and} \ \hat{S}\ket{m}\ket{L} = \ket{m-1}\ket{L}.
\end{equation}

We can deduce a linear operator $\hat{S}$.
\begin{equation}
\hat{S} = \sum_{n=-\infty}^{\infty} \ket{m+1}\bra{m}\otimes\ket{R}\bra{R}+\sum_{n=-\infty}^{\infty} \ket{m-1}\bra{m}\otimes\ket{L}\bra{L}.
\end{equation}

The walker's direction of the walk is decided by the result of the coin operator. The walk consists of applying, at each step, the coin operator, then the shift operator. The combined operation is given~by:
\begin{equation}
\hat{V} = \hat{S}\cdot(\hat{I} \otimes \hat{C}) \label{eqn:walk_coin}.
\end{equation}

This $\hat{V}$ is applied on an initial state multiple times to perform walks with multiple steps.
\begin{equation}
\ket{\psi(t=N)}=\hat{V}^N\ket{\psi(t=0)}.
\end{equation}

The Hadamard coin operator $\hat{H_2}$ can be used to demonstrate the quantum walk. The coin operator $\hat{C}$ in Equation~(\ref{eqn:walk_coin}) is substituted by $\hat{H_2}$.
\begin{equation}
\hat{H_2} = \frac{1}{\sqrt{2}}
\begin{pmatrix}
1&1\\
1&-1\\
\end{pmatrix}.
\end{equation}

One cycle of the quantum walk is completed by applying $\hat{H}\otimes \hat{I}$ followed by $\hat{S}$. This process is performed multiple times without making intermediate measurements. Final measurements are made after a certain number of time steps. The probability distribution generated using a quantum walk behaves differently than a classical random walk. The standard deviation of the classical random walk on a line with N step is known to have a size of $\sqrt{N}$~\cite{kempe2003quantum}; on the other hand, a quantum walk on a line has a standard deviation of the order of N. This indicates that the quantum walk spreads faster than the classical random walk and can result in large speed increases in searching applications.

\subsection{Scattering Quantum Walk: Quantum Walk on Edges}
A different picture of the quantum walk is provided by the scattering model introduced~\cite{feldman2004scattering,feldman2007modifying}. This discrete-time scattering-based quantum walk is also called an edge walk. Unlike the coin model, the interference occurs on edges instead of performing the walk only on vertices. Each vertex works as a scattering center in this model. An input photon amplitude and phase will be controlled by a transmission and reflection coefficient at the scattering center. This model starts with a photon in a state $\ket{m-1,m}$, representing a photon propagating from a vertex location $m-1$ 
 to $m$, hence describing a state on an edge.

In this scattering model, a Michelson interferometer can serve as a vertex with two edges.
One-step propagation starting at a specific edge is given by:
\begin{equation}
\begin{split}
\hat{U}_{Michelson}\ket{m-1,m} \rightarrow \frac{1}{\sqrt{2}}(\ket{m,m+1}+i\ket{m,m-1}).\\
\hat{U}_{Michelson}\ket{m+1,m} \rightarrow \frac{1}{\sqrt{2}}(\ket{m,m-1}+i\ket{m,m+1}).
\end{split}
\label{eqn:michelson}
\end{equation}

\begin{figure}[H]
 \centering
 \begin{subfigure}[t]{0.5\textwidth}
 \centering
 \includegraphics[height=0.8in]{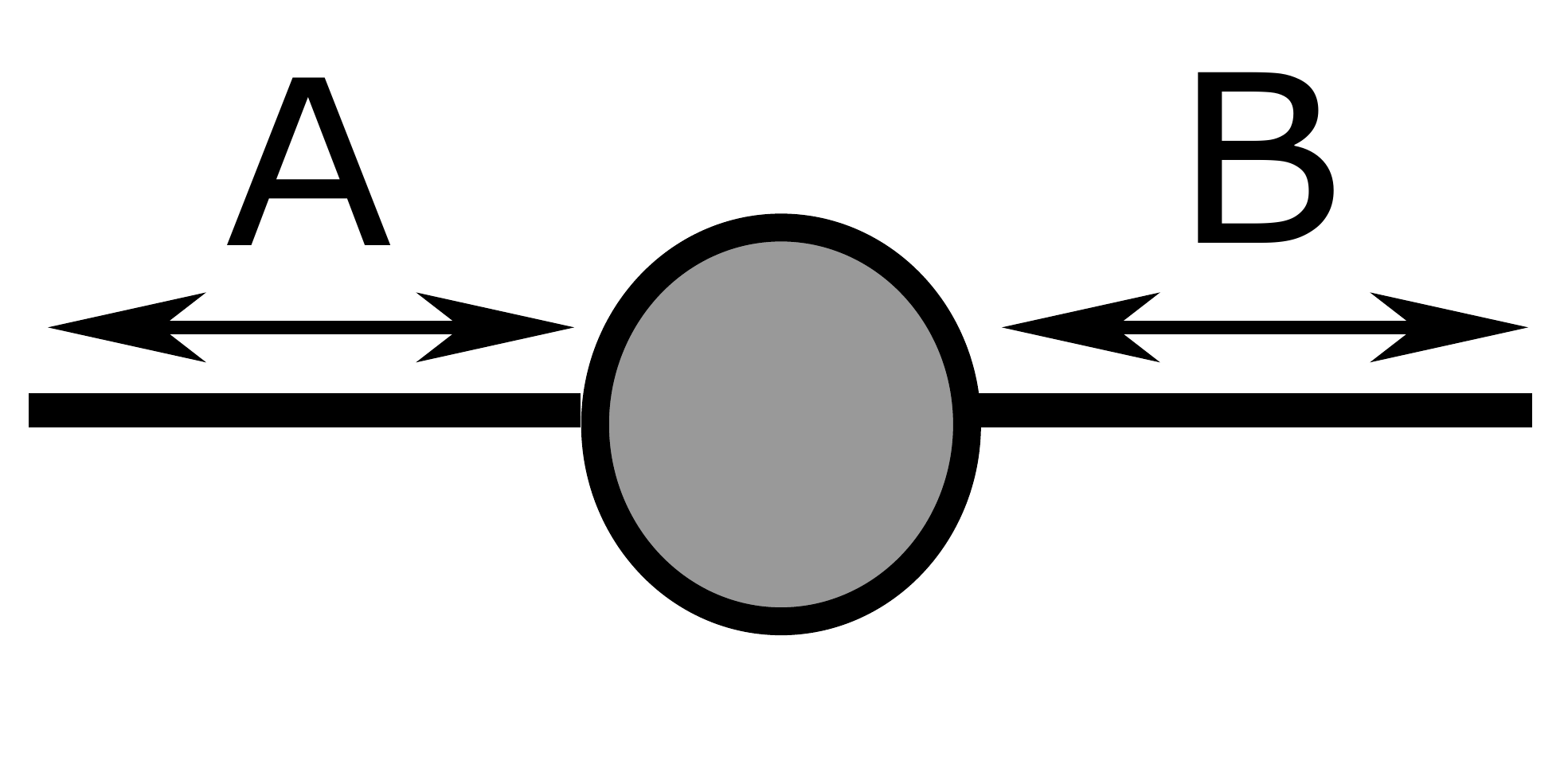}
 \caption{}
 \end{subfigure}%
 ~
 \begin{subfigure}[t]{0.5\textwidth}
 \centering
 \includegraphics[height=0.8in]{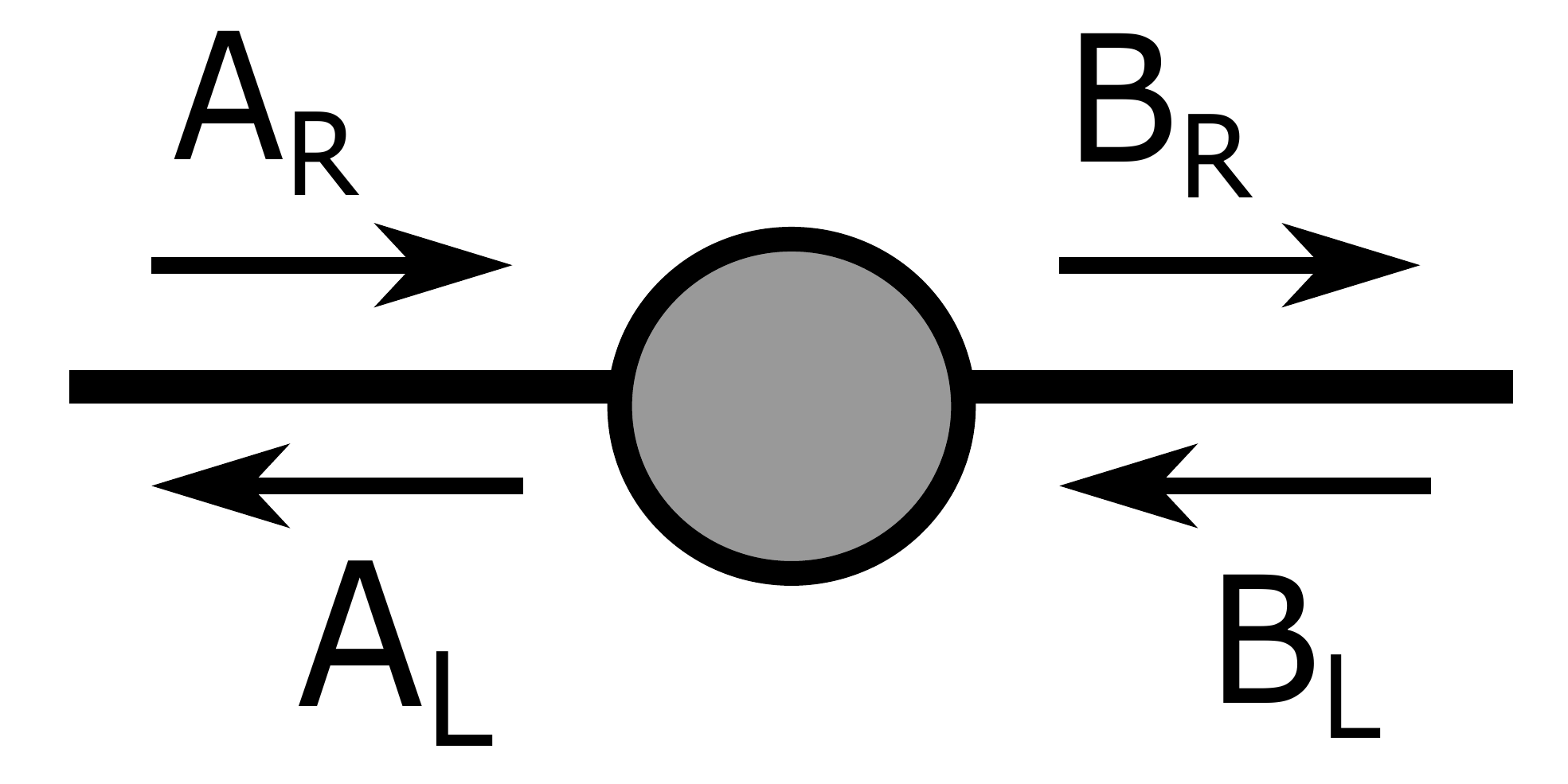}
 \caption{}
 \end{subfigure}
 \caption{(\textbf{a}) Directionally-{insensitive} description for a scattering center. A photon can enter Port A and then leave either Port A or B. (\textbf{b}) Directionally-{sensitive} description for a scattering center. If the photon is initially in the state $A_R$, then the photon will have an amplitude in the $A_L$ direction and the $B_R$ direction. Photon in states $A_R$ and $A_L$ do not interact, so they need to be distinguished when a graph is formed based on scattering~centers. \label{fig:scattering}}
\end{figure}

We will introduce a simplified description for a single scattering center, but we present the full description first. The unitary transformation represented by the Michelson interferometer in Equation~(\ref{eqn:michelson}) and illustrated in Figure~\ref{fig:scattering}{b} can be rewritten using the matrix below:
\begin{equation}
U_{Full} =
\begin{pmatrix}
A_R \rightarrow A_R & A_L \rightarrow A_R & B_R \rightarrow A_R & B_L \rightarrow A_R\\
A_R \rightarrow A_L & A_L \rightarrow A_L & B_R \rightarrow A_L & B_L \rightarrow A_L\\
A_R \rightarrow B_R & A_L \rightarrow B_R & B_R \rightarrow B_R & B_L \rightarrow B_R\\
A_R \rightarrow B_L & A_L \rightarrow B_L & B_R \rightarrow B_L & B_L \rightarrow B_L
\end{pmatrix}
= \frac{1}{\sqrt{2}}
\begin{pmatrix}
0 & i & 1 & 0\\
i & 0 & 0 & 1\\
1 & 0 & 0 & i\\
0 & 1 & i & 0
\end{pmatrix}.
\end{equation}

\begin{equation}
\begin{split}
\hat{U}_{Full}\ket{A_R} \rightarrow \frac{1}{\sqrt{2}}(\ket{B_R}+i\ket{A_L}),\\
\hat{U}_{Full}\ket{B_L} \rightarrow \frac{1}{\sqrt{2}}(\ket{A_L}+i\ket{B_R}).
\end{split}
\end{equation}

$\ket{A_R}$, $\ket{A_L}$, $\ket{B_R}$, and $\ket{B_L}$ correspond to $\ket{m-1,m}$, $\ket{m,m-1}$, $\ket{m,m+1}$, and $\ket{m+1,m}$, respectively.
The propagation direction needs to be distinguished when multiple scattering centers are connected, but a simplified version can be used for a single scattering center. We will use the simplified matrix for a single scattering element in the upcoming sections so that we can directly make a comparison to the coin operators:
\begin{equation}
\hat{U}_{Simplified} =
\begin{pmatrix}
A \rightarrow A & B \rightarrow A\\
A \rightarrow B & B \rightarrow B
\end{pmatrix}
= \frac{1}{\sqrt{2}}
\begin{pmatrix}
i & 1\\
1 & i
\end{pmatrix}
\end{equation}

By repeating this unitary matrix transformation process from an initial state, a walk can be implemented on a line. This scattering model is unitarily equivalent to the coin walk~\cite{venancio2013unveiling,hillery2003quantum}. To see the unitary equivalence between the two models, we define a unitary operator $\hat{E}$.
\begin{equation}
\begin{split}
\hat{E}\ket{m-1,m} = \ket{m}\otimes\ket{R},\\
\hat{E}\ket{m+1,m} = \ket{m}\otimes\ket{L},
\end{split}
\end{equation}
where $\ket{R}$ and $\ket{L}$ are defined in the coin model section. Consider a state evolution by operators $\hat{E}\hat{U}$ with an initial condition $\ket{m-1,m}.$

\begin{equation}
\begin{split}
\hat{U}\ket{m-1,m} = \frac{1}{\sqrt{2}}(\ket{m,m+1}+i\ket{m,m-1}),\\
\hat{E}\hat{U}\ket{m-1,m} = \frac{1}{\sqrt{2}}(\ket{m+1}\otimes\ket{R}+i\ket{m-1}\otimes\ket{L}).
\end{split}
\end{equation}

\begin{equation}
\begin{split}
\hat{V}\hat{E}\ket{m-1,m} = \hat{V}\ket{m}\otimes\ket{R} = \hat{S}\ket{m}\otimes \frac{1}{\sqrt{2}}(\ket{R}+i\ket{L})\\
= \frac{1}{\sqrt{2}}(\ket{m+1}\otimes \ket{R}+i\ket{m-1}\otimes \ket{L}).
\end{split}
\end{equation}

The former represents the edge walk, and the latter represents the coin walk. The outcomes are the same when the evolution operators $\hat{U}$ and $\hat{V}$ are multiplied by the operator $\hat{E}$; therefore, these two formalisms are unitarily equivalent. This unitary equivalence $\hat{E}\hat{U} = \hat{V}\hat{E}$ can be also seen as $\hat{U} = \hat{E}^{\dagger}\hat{V}\hat{E}$. $\hat{U}^n = \hat{E}^{\dagger}\hat{V}^n\hat{E}$ because of unitarity of the operator $\hat{E}$ where n is an integer. This result can be extended to higher dimensional walks. We can find the same equivalence for an initial state $\ket{m+1,m}$.\\

\subsection{Higher Dimensional Coin Operators and Scattering Vertices} \label{sec:coin_operators}
Quantum walks can be extended to higher dimensions by changing the dimension of the operators in the system and attaching additional edges to each vertex. It is possible to introduce scattering centers with different scattering amplitude ratios between output modes using directionally-unbiased devices. We will introduce several different coin operators and corresponding scattering centers in this section. The relationship between the coin model and the scattering model is deduced using an additional unitary operator as discussed in the previous subsection. There are several quantum coin operators with specific characteristics. The Hadamard coin, an unbiased coin, is one example.

The four-dimensional real-valued Hadamard coin $H_4$ is given as an example. This matrix is obtained by taking tensor product of two two by two real Hadamard matrices $H_2$.
\begin{equation}
H_4 = H_2\otimes H_2=\frac{1}{2}
\begin{pmatrix}
1&1&1&1\\
1&-1&1&-1\\
1&1&-1&-1\\
1&-1&-1&1\\
\end{pmatrix}.
\end{equation}

In addition to the Hadamard coin, there are two other major specific coins used in quantum information processing. The first coin is motivated by Grover's search algorithm~\cite{grover1996fast}.
\begin{equation}
C_d =
\begin{pmatrix}
\frac{2}{d}-1 & \frac{2}{d} & \dots & \frac{2}{d}\\
\frac{2}{d} & \frac{2}{d}-1 & \dots & \frac{2}{d}\\
\vdots & \vdots &\ddots& \vdots\\
\frac{2}{d} & \frac{2}{d} & \dots & \frac{2}{d}-1
\end{pmatrix},
\end{equation}
where d is the size of the matrix.

Matrices for $d=3$ and 4 are given.
\begin{equation}
C_3 = \frac{1}{3}
\begin{pmatrix}
-1 &2& 2\\
2 &-1& 2\\
2& 2& -1
\end{pmatrix}
\quad \mbox{ and }\quad
C_4 =\frac{1}{2}
\begin{pmatrix}
-1 &1& 1&1\\
1 &-1& 1&1\\
1& 1& -1&1\\
1&1&1&-1
\end{pmatrix}.
\end{equation}

This coin is biased in amplitudes (except for d = 4), yet symmetric under permutations of matrix labels. Another coin is a discrete Fourier transform (DFT) coin; this coin is unbiased; however, it is not symmetric under permutations. The Fourier transform matrix is given by:
\begin{equation}
U_{Fourier} = \frac{1}{\sqrt{d}}
\begin{pmatrix}
1 & 1 & 1 & \dots & 1\\
1 & \omega & \omega^2 & \dots & \omega^{d-1}\\
1 & \omega^2& \omega^4 & \dots & \omega^{2(d-1)}\\
\vdots & \vdots&\vdots &\ddots &\vdots\\
1 & \omega^{(d-1)}& \omega^{2(d-1)}& \dots &\omega^{(d-1)(d-1)}
\end{pmatrix},
\end{equation}
where $\omega = e^{-\frac{2\pi i}{d}}$.

Three-dimensional and four-dimensional Fourier coins are given by:
\begin{equation}
U_{Fourier} = \frac{1}{\sqrt{3}}
\begin{pmatrix}
1 & 1 & 1\\
1 & \omega_3 & \omega_3^2 \\
1 & \omega_3^2& \omega_3\\
\end{pmatrix},
\end{equation}
where $\omega_3 = e^{-\frac{2\pi i}{3}}$.

\begin{equation}
U_{Fourier} = \frac{1}{2}
\begin{pmatrix}
1 & 1 & 1& 1\\
1 & \omega_4 & \omega_4^2 & \omega_4^{3}\\
1 & \omega_4^2& \omega_4^4 & \omega_4^{6}\\
1 & \omega_4^{3}& \omega_4^{6} &\omega_4^{9}
\end{pmatrix}
=\frac{1}{2}
\begin{pmatrix}
1 &1& 1&1\\
1 &i& 1&-i\\
1& -1& 1&-1\\
1&-i&-1&i
\end{pmatrix},
\end{equation}
where $\omega_4 = e^{-\frac{2\pi i}{4}}$.

\subsection{Equivalence between Higher Dimensional Coin Walk and Scattering Quantum Walk}
The coin walk and the scattering walk were introduced in the previous subsections, as well as higher dimensional coin operators. It is possible to give a unitary equivalence relation between the two walks in higher dimensions as well. Consider a quantum walk on a 2D rectangular lattice. The center of the grid is given by coordinate ($m,n$). A photon on one of the edges around that grid is defined as $\ket{m,m,n-1,n}$. This state is read as a photon propagation from a vertex location ($m,n-1$) to ($m,n$). The unitary operator for one propagation step of an edge is defined as:
\begin{equation}
\begin{split}
\hat{U}\ket{m-1,m,n,n} \rightarrow \frac{1}{2}(\ket{m,m-1,n,n}+\ket{m,m,n,n+1}-\ket{m,m+1,n,n}+\ket{m,m,n,n-1}),\\
\hat{U}\ket{m,m,n+1,n} \rightarrow \frac{1}{2}(\ket{m,m-1,n,n}+\ket{m,m,n,n+1}+\ket{m,m+1,n,n}-\ket{m,m,n,n-1}),\\
\hat{U}\ket{m+1,m,n,n} \rightarrow \frac{1}{2}(-\ket{m,m-1,n,n}+\ket{m,m,n,n+1}+\ket{m,m+1,n,n}+\ket{m,m,n,n-1}),\\
\hat{U}\ket{m,m,n-1,n} \rightarrow \frac{1}{2}(\ket{m,m-1,n,n}-\ket{m,m,n,n+1}+\ket{m,m+1,n,n}+\ket{m,m,n,n-1}).
\end{split}
\end{equation}

The corresponding coin operator of the coin walk is given by:
\begin{equation}
\hat{C} = \frac{1}{2}
\begin{pmatrix}
-1 & 1 & 1 & 1\\
1 & -1 & 1 & 1\\
1 & 1 & -1 & 1\\
1 & 1 & 1 & -1
\end{pmatrix}.
\end{equation}

The operator transforms an initial coin state into a superposition state:
\begin{equation}
\hat{C}\ket{L} = \frac{1}{2}(-\ket{L}+\ket{U}+\ket{R}+\ket{D}),
\end{equation}
where $\ket{L} = (1, 0, 0, 0)^T,\ket{U} = (0, 1, 0, 0)^T,\ket{R} = (0, 0, 1, 0)^T,\ket{D} = (0, 0, 0, 1)^T.$ A new shift operator is defined as follows:
\begin{equation}
\begin{split}
\hat{S} &= \sum_{m} \sum_{n}(\ket{m,n+1}\bra{m,n}\otimes\ket{U}\bra{U}+\ket{m,n-1}\bra{m,n}\otimes\ket{D}\bra{D}\\
&+\ket{m+1,n}\bra{m,n}\otimes\ket{R}\bra{R}+\ket{m-1,n}\bra{m,n}\otimes\ket{L}\bra{L}).
\end{split}
\end{equation}

One step of the coin walk is given by:
\begin{equation}
\hat{V} = \hat{S}(\hat{I}\otimes \hat{C}).
\end{equation}

We wish to find equivalence between the edge walk and the coin walk by finding a unitary operator $\hat{E}$. Define an operator $\hat{E}$ transforming edge states into vertex states.
\begin{equation}
\begin{split}
\hat{E}\ket{m-1,m,n,n} = \ket{m,n}\otimes\ket{R}\\
\hat{E}\ket{m,m,n+1,n} = \ket{m,n}\otimes\ket{D}\\
\hat{E}\ket{m+1,m,n,n} = \ket{m,n}\otimes\ket{L}\\
\hat{E}\ket{m,m,n-1,n} = \ket{m,n}\otimes\ket{U}.
\end{split}
\end{equation}

Consider two cases for the coin-based walk and edge walk starting with an initial state $\ket{m,m,n-1,n}$.

Unitary transformation of the coin-based walk:
\begin{equation}
\hat{E}\ket{m,m,n-1,n} = \ket{m,n}\otimes\ket{U}.
\end{equation}
\vspace{-32pt}

\begin{myequation3}
\begin{split}
\hat{V}\hat{E}\ket{m,m,n-1,n} &= S\ket{m,n}\otimes(\frac{1}{2}(\ket{L}-\ket{U}+\ket{R}+\ket{D})\\
&=S\frac{1}{2}(\ket{m,n}\otimes\ket{L}-\ket{m,n}\otimes\ket{U}+\ket{m,n}\otimes\ket{R}+\ket{m,n}\otimes\ket{D})\\
&=\frac{1}{2}(\ket{m-1,n}\otimes\ket{L}-\ket{m,n+1}\otimes\ket{U}+\ket{m+1,n}\otimes\ket{R}+\ket{m,n-1}\otimes\ket{D}).
\end{split}
\end{myequation3}

Unitary transformation of the edge-based walk:
\begin{equation}
\begin{split}
\hat{U}\ket{m,m,n-1,n} &= \frac{1}{2}(\ket{m,m-1,n,n}-\ket{m,m,n,n+1}+\ket{m,m+1,n,n}+\ket{m,m,n,n-1}).
\end{split}
\end{equation}
\vspace{-22pt}

\begin{myequation4}
\begin{split}
\hat{E}\hat{U}\ket{m,m,n-1,n} &=\frac{1}{2}(\ket{m-1,n}\otimes\ket{L}-\ket{m,n+1}\otimes\ket{U}+\ket{m+1,n}\otimes\ket{R}+\ket{m,n-1}\otimes\ket{D}).
\end{split}
\end{myequation4}

$\hat{V}\hat{E}$ and $\hat{E}\hat{U}$ both transform the initial state into the same state. Therefore, the outcomes are equivalent, and the coin walk and the scattering walk are unitarily equivalent. We went through a specific equivalence, which is the quantum walk generalized equivalence between two models, as found elsewhere~\cite{venancio2013unveiling,feldman2004scattering}.

\subsection{Examples of Multi-Dimensional Quantum Walks on Graphs}
\begin{figure}[H]
 \centering
 \begin{subfigure}[t]{0.5\textwidth}
 \centering
 \includegraphics[height=1.5in]{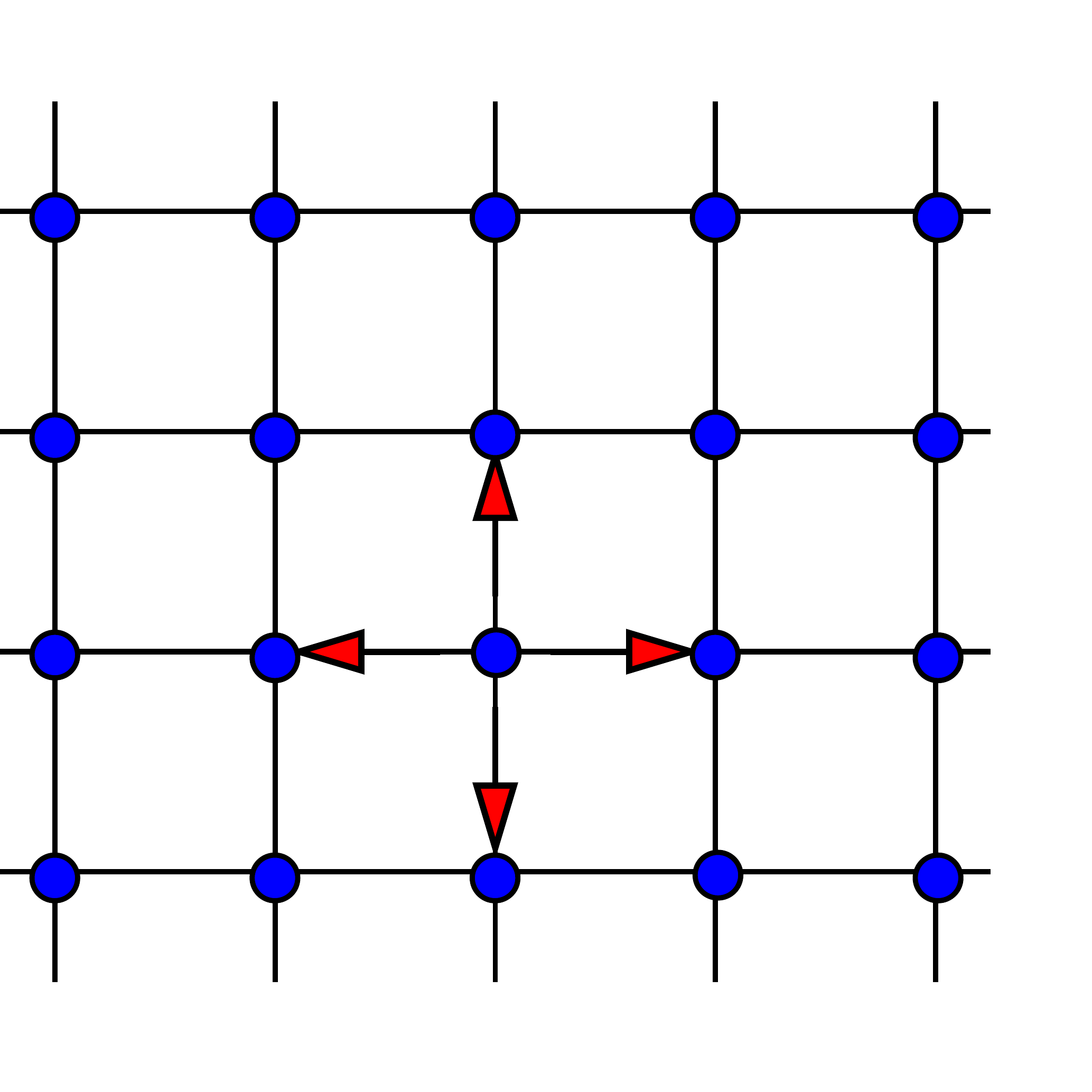}
 \caption{}\label{fig:glued_tree}
 \end{subfigure}%
 ~
 \begin{subfigure}[t]{0.5\textwidth}
 \centering
 \includegraphics[height=1.5in]{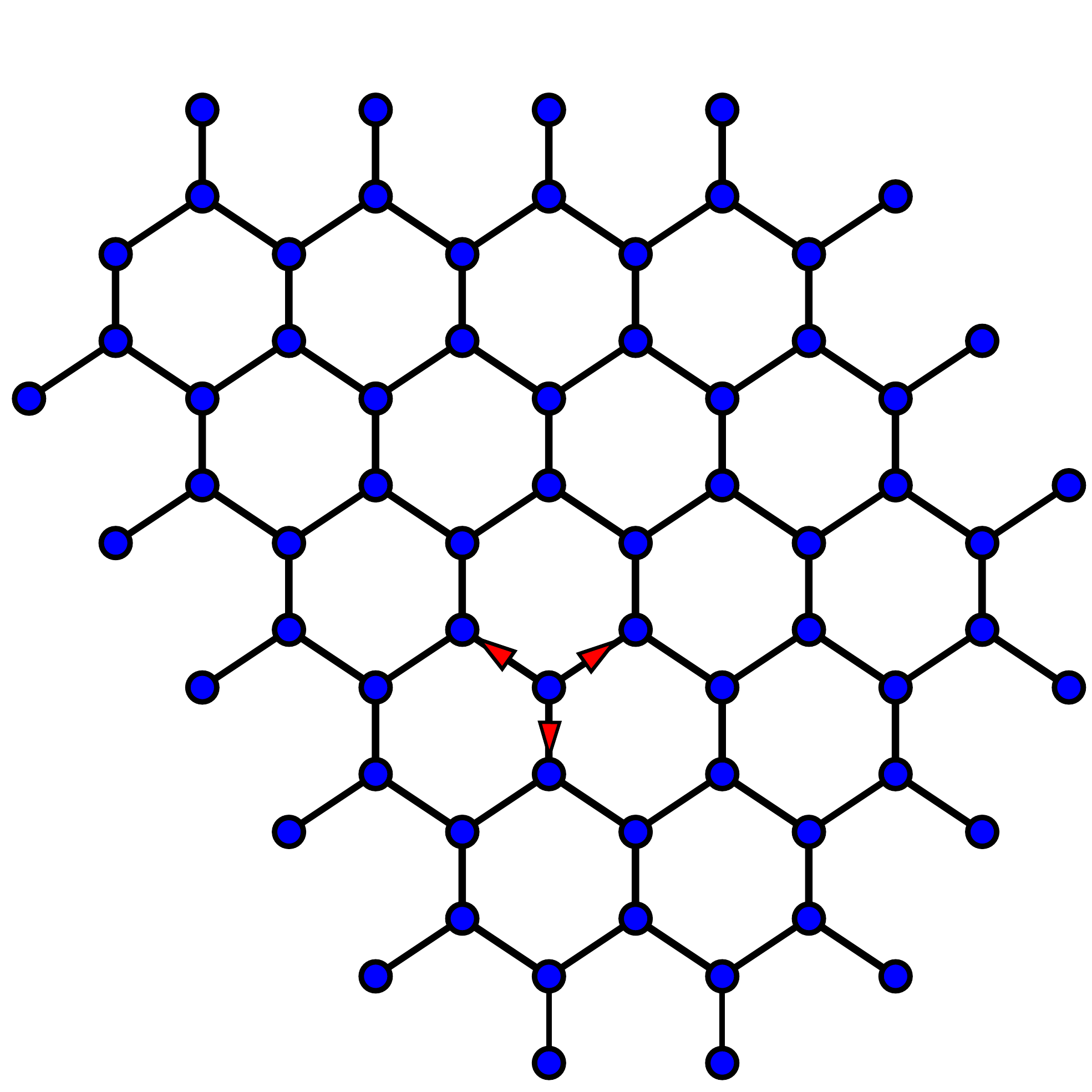}
 \caption{}\label{fig:4d_hyper}
 \end{subfigure}
 \caption{(\textbf{a}) Rectangular lattice. This structure has four edges for every vertex. (\textbf{b})~Hexagonal lattice. This structure has three edges for every vertex. \label{fig:spatial_search}}
\end{figure}

The two types of quantum walks, the coin quantum walk and the scattering (edge) quantum walk, are both performed on graphs. Graphs with nodes implemented by higher dimensional coins are applicable to algorithm development. For example, the Grover search algorithm, when implemented via quantum walks on certain graphs with a superposition initial state, demonstrates significant speedup over classical algorithms. Many quantum walk applications are based on undirected graphs, meaning a walker can travel forward and backward in the system. Directionally-unbiased linear-optical devices possess reversibility and therefore can implement such undirected graphs.
It is shown that a spatial search performed on a 2D lattice is faster than similar classical algorithms~\cite{benioff2000space,aaronson2003quantum,childs2004spatial,tulsi2008faster,abal2010spatial,abal2012spatial}. To observe the spatial search on a 2D lattice, the scattering centers have transmission and reflection coefficients equal to the Grover coin setting. One node in a graph is ``marked'' by introducing a different matrix on one specific scattering center in the lattice. Localization occurs on edges around the marked scattering center when the superposition state is sent in the system as an initial state. The graph geometry can be configured using directionally-unbiased devices. The rectangular lattice illustrated in Figure~\ref{fig:spatial_search}{a} would require four-port devices. Similarly, the hexagonal lattice illustrated in Figure~\ref{fig:spatial_search}{b} would require three-port devices. These optical quantum walk implementations through these optical devices are advantageous because of their amplitude tunability. The graphs can be implemented with the same Grover coin matrix setting throughout the graph vertices initially, then a marked coin can be introduced by tuning one of the vertices into a different coin. A quantum walk search can find the marked point faster than any classical search algorithms.

\begin{figure}[H]
 \centering
 \begin{subfigure}[t]{0.5\textwidth}
 \centering
 \includegraphics[height=1.5in]{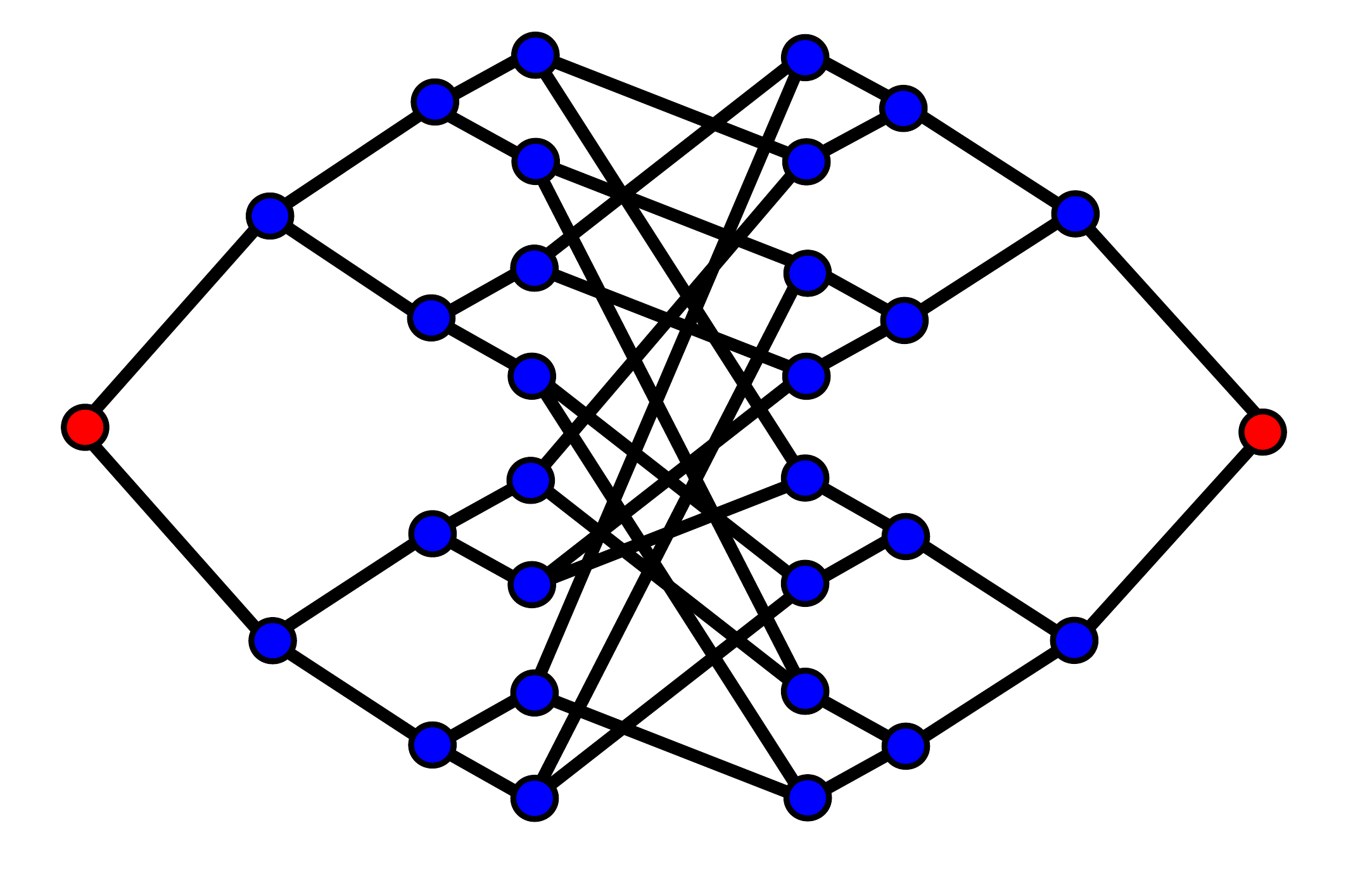}
 \caption{}\label{fig:glued_tree}
 \end{subfigure}%
 ~
 \begin{subfigure}[t]{0.5\textwidth}
 \centering
 \includegraphics[height=1.5in]{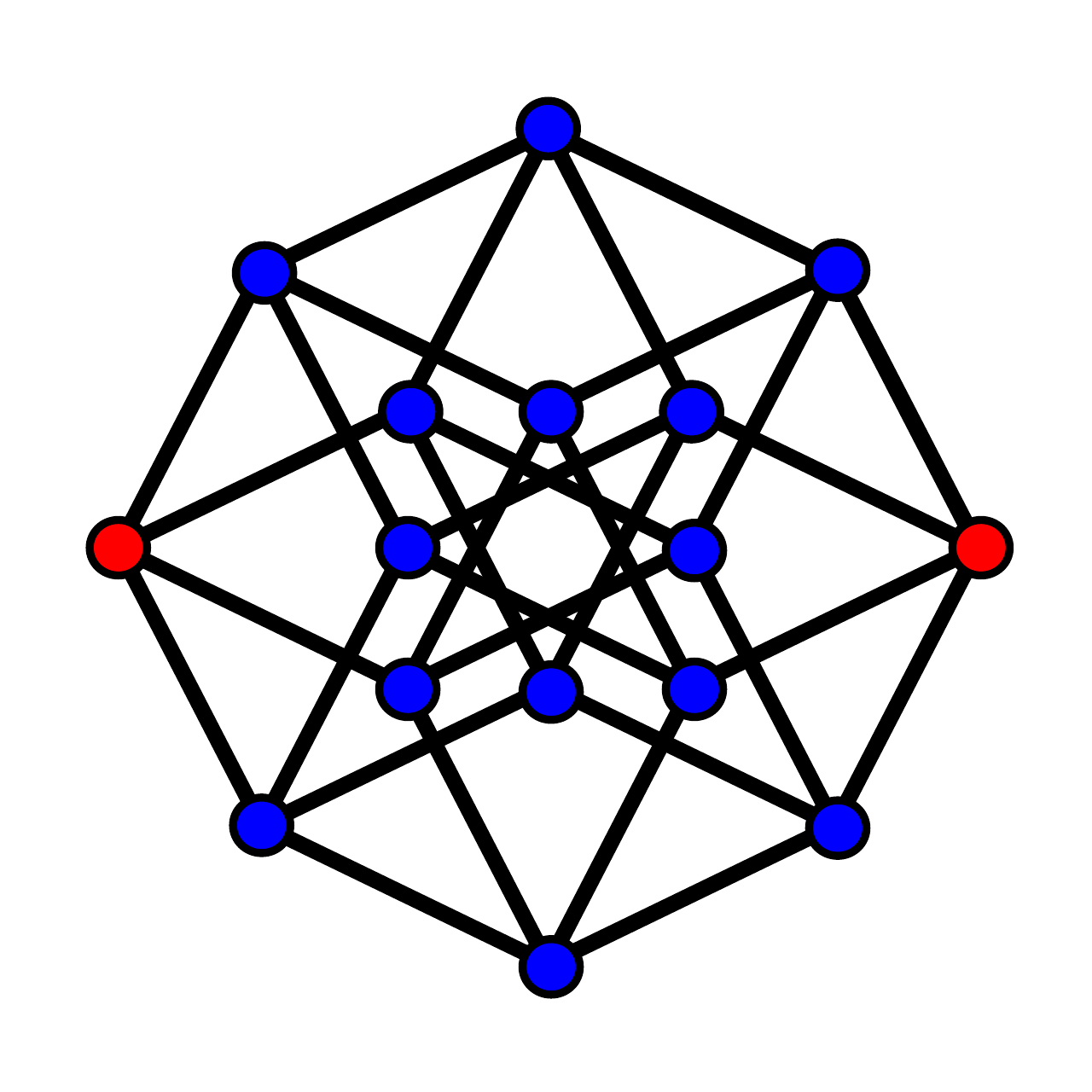}
 \caption{}\label{fig:4d_hyper}
 \end{subfigure}
 \caption{(\textbf{a}) Glued tree. Two trees are glued in a random manner in the middle part of the graph. A photon's hitting time from one red circle to the other red circle is shorter than the classical walk on this graph. (\textbf{b})~Four-dimensional hypercube. Every vertex has four edges. A photon's hitting time from one red circle to the other red circle is shorter than the classical walk on this graph. \label{fig:hitting_time}}
\end{figure}

Other graph structures can be considered using directionally-unbiased devices. Quantum walks on a glued tree have been investigated theoretically and experimentally~\cite{kempe2005discrete,tang2018experimental}. Hitting time, the time required to reach one point to another point on a graph, is commonly used to evaluate propagation speed on a specific graph. A quantum walk on a glued tree with three nodes gives exponential speedup when the three-dimensional Grover coin is used at the nodes~\cite{tregenna2003controlling}. We can form hypercubes using unbiased multiports, and it has been shown that the quantum walk hitting time is shorter than the classical walk case~\cite{krovi2006hitting}. A walker starts on the left side of the graph, and the walker tries to reach the other end in a short amount of time. As indicated in Figure~\ref{fig:hitting_time}{a}, the randomly-connected middle part in the glued tree complicates the path finding procedure to reach the other end. The speedup applies in the case of the hypercube as well. A four-dimensional hypercube is illustrated in Figure~\ref{fig:hitting_time}{b} as an example. Classical algorithms cannot perform this search efficiently. On the other hand, quantum walks can perform exponentially faster than any classical algorithms to find the other end of the path.

\section{Specific Transfer Matrix Examples Using Reversible Linear-Optical Devices}

We consider several specific experimental configurations for an efficient realization of quantum walks in higher dimensions using linear optical devices and exploiting the very important feature of optical reversibility. We look into specific phase values and corresponding transfer matrices using formalisms covered in previous sections. The focus here is on realization of the Fourier coin and the Grover coin using directionally-unbiased linear-optical multiports, and reversible optical tritter and quarter configurations.

\subsection{The Fourier Coin Realization}

A three-dimensional Fourier coin has the form (see Section~\ref{sec:coin_operators}):
\begin{equation}
U_{Fourier} = \frac{1}{\sqrt{3}}
\begin{pmatrix}
1 & 1 & 1\\
1 & \omega_3 & \omega_3^2 \\
1 & \omega_3^2& \omega_3\\
\end{pmatrix},
\end{equation}

\noindent where $\omega_3 = e^{-\frac{2\pi i}{3}}$.

\begin{figure}[H]
\includegraphics[scale=0.25]{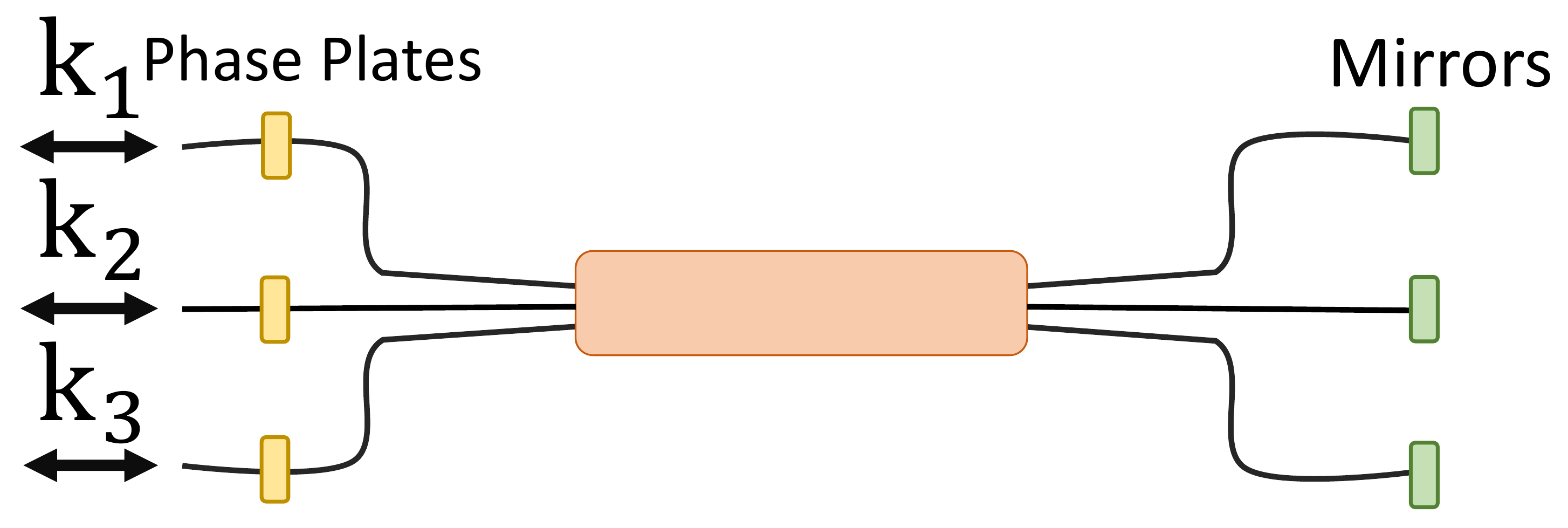}
\centering
\caption{Reversible tritter with phase shifters. The yellow squares are phase shifters, and the green squares are mirrors. \label{fig:reversible_tritter_phase}}
\end{figure}

This matrix can be generated with a reversible tritter containing phase shifters at all three ports, and directionally-unbiased linear-optical three-ports can perform the same job as well. For a reversible system, an input photon experiences the same phase shifts twice from the same phase shifters. $U_{phasein}$ and $U_{phaseout}$ would take care of such phase shifts. Figure~\ref{fig:reversible_tritter_phase} is a reversible tritter with phase shifters. Multiport designs are introduced in Section \ref{multiports_sub}.

The matrix generation is performed by multiplying $U_{phasein}$, $U_{device}$, and $U_{phaseout}$ in sequence. $U_{device}$ can be any reversible optical device. The Fourier coin using the optical tritter is realized by the following procedure.
\begin{equation}
U_{Fourier} = U_{phaseout}U_{Tritter}U_{phasein}.
\end{equation}
\begin{equation}
\begin{split}
U_{phasein} = U_{phaseout}=
\begin{pmatrix}
e^{i\phi_a} & 0 & 0\\
0 & e^{i\phi_b} & 0\\
0 & 0 & e^{i\phi_c}
\end{pmatrix},
\end{split}
\end{equation}

\begin{equation}
\begin{split}
U_{Fourier} &=
\begin{pmatrix}
e^{-i\frac{\pi}{3}} & 0 & 0\\
0 & e^{i\frac{\pi}{3}} & 0\\
0 & 0 & e^{i\frac{\pi}{3}}
\end{pmatrix}
\frac{i}{\sqrt{3}}
\begin{pmatrix}
e^{i\frac{2\pi}{3}} & 1 & 1\\
1 & e^{i\frac{2\pi}{3}} & 1\\
1 & 1 & e^{i\frac{2\pi}{3}}
\end{pmatrix}
\begin{pmatrix}
e^{-i\frac{\pi}{3}} & 0 & 0\\
0 & e^{i\frac{\pi}{3}} & 0\\
0 & 0 & e^{i\frac{\pi}{3}}
\end{pmatrix}\\
&=\frac{i}{\sqrt{3}}
\begin{pmatrix}
1 & 1 & 1\\
1 & e^{-i\frac{4\pi}{3}} & e^{-i\frac{2\pi}{3}}\\
1 & e^{-i\frac{2\pi}{3}} & e^{-i\frac{4\pi}{3}}
\end{pmatrix},
\end{split}
\end{equation}

\noindent where $\phi_a, \phi_b,\phi_c$ are the phase shifts from phase shifters at the entrance ports. The reversible optical tritter and quarter can realize the Fourier coin when phases are set at specific values:
\begin{equation}
\begin{split}
&(\phi_A,\phi_B,\phi_C,\kappa z,\phi_a,\phi_b,\phi_c)= (\frac{10\pi}{9},\frac{10\pi}{9},\frac{10\pi}{9},\frac{10\pi}{9},-\frac{\pi}{3},\frac{\pi}{3},\frac{\pi}{3}),
\end{split}
\end{equation}

\noindent where $\phi_A,\phi_B,\phi_C$, and $\kappa z$ are phase values from mirror units of the reversible tritter and propagation distance of the coupling region.

Similarly, the unbiased three-port operation can realize the same matrix with the settings:
\begin{equation}
U_{Fourier} = U_{phaseout}U_{Three-port}U_{phasein}.
\end{equation}

\begin{equation}
\begin{split}
U_{Fourier} &=
\begin{pmatrix}
e^{i\frac{\pi}{3}} & 0 & 0\\
0 & e^{-i\frac{\pi}{3}} & 0\\
0 & 0 & e^{-i\frac{\pi}{3}}
\end{pmatrix}
\frac{1}{\sqrt{3}}e^{i\frac{2\pi}{3}}
\begin{pmatrix}
e^{-i\frac{2\pi}{3}} & 1 & 1\\
1 & e^{-i\frac{2\pi}{3}} & 1\\
1 & 1 & e^{-i\frac{2\pi}{3}}
\end{pmatrix}
\begin{pmatrix}
e^{i\frac{\pi}{3}} & 0 & 0\\
0 & e^{-i\frac{\pi}{3}} & 0\\
0 & 0 & e^{-i\frac{\pi}{3}}
\end{pmatrix}\\
&=\frac{1}{\sqrt{3}}e^{i\frac{2\pi}{3}}
\begin{pmatrix}
1 & 1 & 1\\
1 & e^{-i\frac{4\pi}{3}} & e^{-i\frac{2\pi}{3}}\\
1 & e^{-i\frac{2\pi}{3}} & e^{-i\frac{4\pi}{3}}
\end{pmatrix},
\end{split}
\end{equation}
when:
\begin{equation}
\begin{split}
(\phi_A,\phi_B,\phi_C,\phi_a,\phi_b,\phi_c)= (\frac{\pi}{6},\frac{\pi}{6},\frac{\pi}{6}, \frac{\pi}{3}, -\frac{\pi}{3}, -\frac{\pi}{3}),
\end{split}
\end{equation}
where $\phi_A,\phi_B$, and $\phi_C$ are the phase values for the directionally-unbiased three-port operation.

\begin{figure}[H]
\includegraphics[scale=0.25]{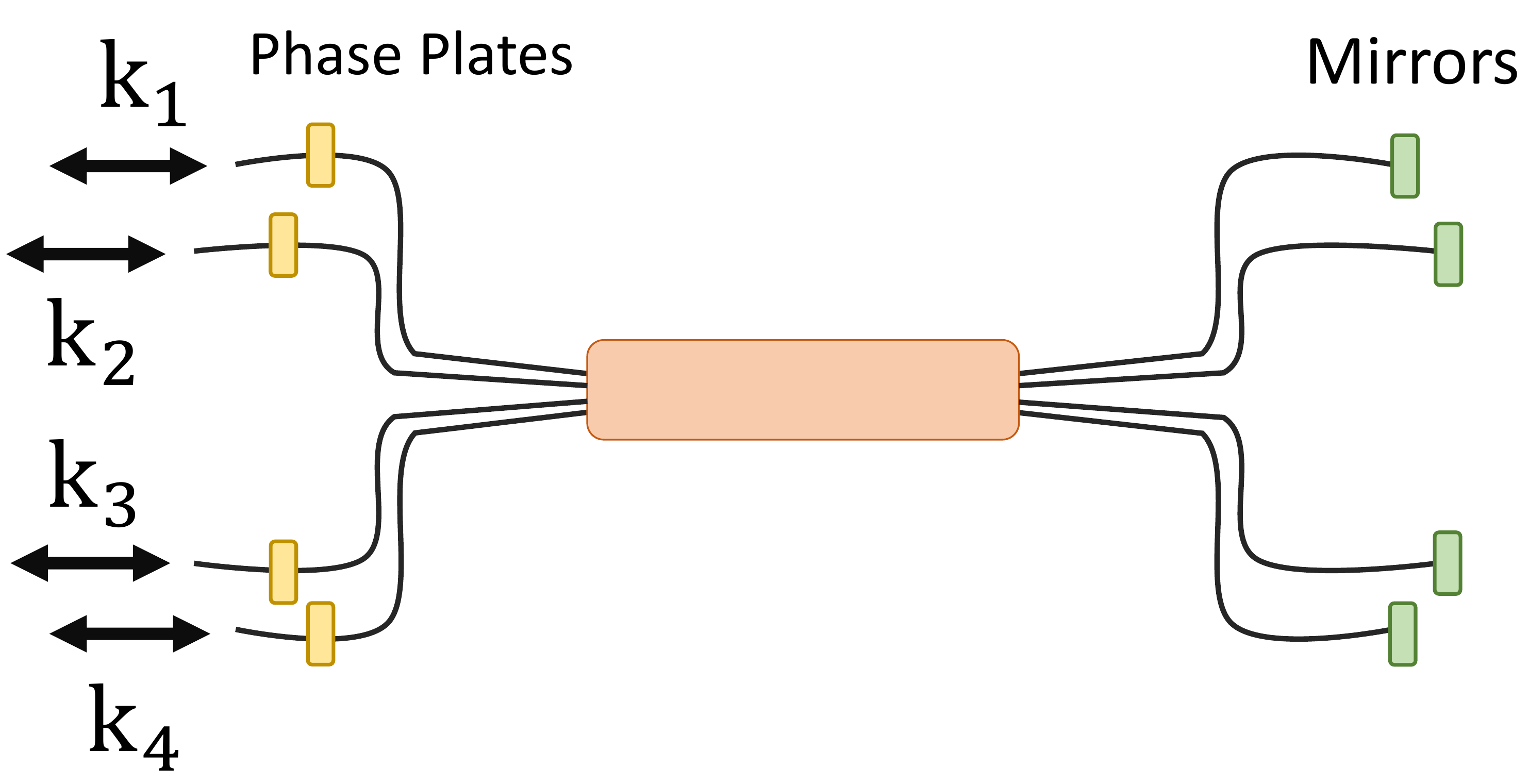}
\centering
\caption{Reversible quarter with phase shifters. The yellow squares are phase shifters, and the green squares are mirrors. \label{fig:reversible_quarter_phase}}
\end{figure}

Four-dimensional matrices are generated using the same methods. The reversible quarter with phase shifters is given in Figure~\ref{fig:reversible_quarter_phase}. Multiport designs are introduced in Section \ref{multiports_sub}.

The four-dimensional Fourier coin has the form of:
\begin{equation}
U_{Fourier} = \frac{1}{2}
\begin{pmatrix}
1 & 1 & 1& 1\\
1 & \omega_4 & \omega_4^2 & \omega_4^{3}\\
1 & \omega_4^2& \omega_4^4 & \omega_4^{6}\\
1 & \omega_4^{3}& \omega_4^{6} &\omega_4^{9}
\end{pmatrix},
\end{equation}
where $\omega_4 = e^{-\frac{2\pi i}{4}}$.\\
\begin{equation}
U_{Fourier} = U_{phaseout}U_{Quarter}U_{phasein} ,
\end{equation}
\begin{equation}
\begin{split}
U_{phasein} = U_{phaseout}=
\begin{pmatrix}
e^{i\phi_a} & 0 & 0&0\\
0 & e^{i\phi_b} & 0&0\\
0 & 0 & e^{i\phi_c}&0\\
0&0&0& e^{i\phi_d}
\end{pmatrix}.
\end{split}
\end{equation}

The reversible quarter can realize the Fourier coin as well with the phase settings equal to:
\begin{equation}
\begin{split}
&(\phi_A,\phi_B,\phi_C,\phi_D,\kappa z_1,\kappa z_2,\phi_a,\phi_b,\phi_c,\phi_d) = (\pi,\frac{\pi}{4},\pi,\frac{5\pi}{4},\frac{7\pi}{4},\frac{7\pi}{8},-\frac{\pi}{4},-\frac{\pi}{2},-\frac{\pi}{4},\frac{\pi}{2}),
\end{split}
\end{equation}
where $\phi_A,\phi_B,\phi_C,\phi_D$, and $\kappa z_1,\kappa z_2$ are phase values from the mirror units of the reversible tritter and propagation distance of the coupling region.

Four ports:
\begin{equation}
U_{Fourier} = U_{phaseout}U_{fourport}U_{phasein},
\end{equation}
when:
\begin{equation}
\begin{split}
(\phi_A,\phi_B,\phi_C,\phi_D,\phi_a,\phi_b,\phi_c,\phi_d)= (0,\frac{\pi}{2},0, \frac{\pi}{2},-\frac{\pi}{4},-\frac{\pi}{4},\frac{3\pi}{4},-\frac{\pi}{4}),
\end{split}
\end{equation}

\noindent where $\phi_A,\phi_B,\phi_C,\phi_D$ are the phase settings of the four-port operation.
\subsection{The Grover Coin Realization}
The Grover coin is realized using reversible designs as well. The procedure is identical to the Fourier coin case.
The three-dimensional Grover coin takes the form of:
\begin{equation}
C_3 = \frac{1}{3}
\begin{pmatrix}
 -1 & 2 & 2\\
 2 & -1 & 2\\
 2 & 2 & -1
 \end{pmatrix}.
\end{equation}

This can be realized using a reversible tritter with phase settings equal to:
\begin{equation}
\begin{split}
&(\phi_A,\phi_B,\phi_C,\kappa z) = (\frac{11\pi}{6},\frac{11\pi}{6},\frac{11\pi}{6},\frac{11\pi}{6}),
\end{split}
\end{equation}
or using an unbiased three-port with settings:
\begin{equation}
\begin{split}
&(\phi_A,\phi_B,\phi_C) = (\frac{3\pi}{2},\frac{3\pi}{2},\frac{3\pi}{2}).
\end{split}
\end{equation}

The four-dimensional Grover coin operator is given by:
\begin{equation}
C_4 =\frac{1}{2}
\begin{pmatrix}
-1 &1& 1&1\\
1 &-1& 1&1\\
1& 1& -1&1\\
1&1&1&-1
\end{pmatrix},
\end{equation}

\noindent which can be realized with a reversible quarter,
\begin{equation}
U_{Grover} = U_{phaseout}U_{Quarter}U_{phasein}.
\end{equation}

The phase settings for this Grover coin realization is done by:
\begin{equation}
\begin{split}
&(\phi_A,\phi_B,\phi_C,\phi_D,\kappa z_1,\kappa z_2) = (0,0,0,0,\frac{\pi}{8},\frac{\pi}{8}).
\end{split}
\end{equation}

Similarly for the four ports:
\begin{equation}
U_{Grover} = U_{phaseout}U_{Four-ports}U_{phasein} ,
\end{equation}
\begin{equation}
\begin{split}
&(\phi_A,\phi_B,\phi_C,\phi_D)= (\frac{3\pi}{2},\frac{3\pi}{2},\frac{3\pi}{2},\frac{3\pi}{2}).
\end{split}
\end{equation}

\section{Comparison between Directional- and Directionally-Unbiased Devices}
Directional devices (the Reck and Clements decomposition model) can produce any unitary matrices U(N). However, when reversibility is introduced in the system (reversible tritters and directionally-unbiased linear-optical multiports), it imposes symmetry or the self-transpose property $U_{i,j} = U_{j,i}$, where i,j are matrix indices. Hence, reversible designs only produce the subset of {symmetric} unitary matrices. As an example, Equation~(\ref{eqn:switch}) is a unitary matrix, but it is not a self-transpose~matrix. 
\begin{equation}
U =
\begin{pmatrix}
0 & 1 & 0\\
0 & 0 & 1\\
1 & 0 & 0
\end{pmatrix}.
\label{eqn:switch}
\end{equation}

Nevertheless, this reversible design can produce important coins, such as Grover and Fourier coins, for quantum walks. The properties of each directional- and directionally-unbiased design are compared in Table~\ref{tab:table_comp}. In this table, directionally-unbiased three-port and four-port operations are denoted as 3-port and 4-port; Reversible tritter and quarters are denoted as Rev Tritter and Rev Quarter; 3-port Reck, 4-port Reck, and 4-port Clements represent the directional three-port Reck model, the directional four-port Reck model, and the directional four-port Clements model, respectively. In the table, we consider several different conditions for comparing different optical devices: the numbers of beam splitters used to form a device, the coherence length requirement to generate the final unitary matrix, and dense unitary matrix generation. The Grover and the Fourier coin generation are considered as indicated in the Conditions column in Table \ref{tab:table_comp}. The number of beam splitters varies depending on the design. Directionally-unbiased linear-optical multiports require the fewest beam splitters among all the devices in the table. A directionally-unbiased N-port requires N beam splitters, while other devices with N input and N output ports would require N(N + 1) beam splitters. The multiport device demands a long coherence length because different photons traveling paths with different travel lengths need to be coherently summed. In the reversible designs, Reck decomposition, and the Clements decomposition model, the input photon does not need long coherence lengths because the devices consist of balanced interferometers. All the devices listed in the table can realize both the Grover and the Fourier coins.

\begin{table}[H]
\captionsetup{font = {footnotesize}}
\caption{Comparison between reversible and non-reversible designs. In the table, 3-port and 4-port represent directionally-unbiased linear-optical multiports; Rev Tritter and Rev Quarter represent reversible tritter and reversible quarter; 3-Reck, 4-Reck and 4-Clements represent the three-port and four-port Reck decomposition models and the four-port Clements decomposition model.} \label{tab:table_comp}
\centering
 \begin{tabular}{cccccccc}
\toprule
 {\bf Conditions} & {\bf 3-port}& {\bf 4-port}&{\bf Rev Tritter}&{\bf Rev Quarter}& {\bf 3-Reck} &{\bf 4-Reck}& {\bf 4-Clements}\\
 \midrule
 \# of Beam splitters &3 & 4 &- & - & 12 & 20 & 20\\
\midrule
 Coherence Length& Long &Long&Short&Short&Short& Short & Short \\
\midrule
 U Generation & \xmark & \xmark & \xmark& \xmark&\cmark & \cmark & \cmark \\
\midrule
 Grover Coin & \cmark & \cmark &\cmark & \cmark &\cmark & \cmark &\cmark \\
\midrule
 Fourier Coin & \cmark & \cmark &\cmark & \cmark &\cmark & \cmark &\cmark\\
 \bottomrule

 \end{tabular}
\end{table}

\section{Summary}
Directional optical designs and directionally-unbiased linear-optical designs were reviewed and investigated closely. We studied the use of directionally-unbiased linear-optical designs in quantum walk applications. This directionally-unbiased system allowed us to generate reversible graphs with vertices having multiple edges. This flexibility in graph generation can be useful in quantum walk-based algorithmic speedup as briefly mentioned in the introduction part of section~\ref{quantum_walks} 
. Previously introduced directionally-unbiased designs~\cite{simon2016group,osawa2018experimental}, reversible optical tritter and quarter, can work as scattering centers having three and four edges for quantum walk applications. These designs are advantageous because of the implementation cost reduction by removing directional bias in the system. We focused on Grover and Fourier coin implementations, which are important matrices in quantum information processing. Grover and Fourier matrices can be realized when all the phases are at proper settings. Any unitary matrices can be realized using directional devices; however, this is not the case for the directionally-unbiased design. Directionally-unbiased designs cannot realize non-self-transpose unitary matrices. We have focused on quantum walk and search applications, but similar designs can be applied in other applications. For example, Hamiltonian simulations and topological phase simulations are possible immediate applications of directionally-unbiased designs~\cite{simon2017quantum1,simon2017quantum2,simon2018joint}.
\vspace{6pt}



\authorcontributions{Conceptualization, S.O., D.S. and A.S.; software, S.O.; writing—original draft preparation, S.O.; writing—review and editing, D.S. and A.S.; supervision, D.S. and A.S.; funding acquisition, A.S.}
\funding{This research was funded by the National Science Foundation,  Emerging Frontiers in Research and Innovation - Advancing Communication Quantum Information Research in Engineering (EFRI-ACQUIRE) grant number ECCS-1640968, the Air Force Office of Scientific Research (AFOSR) grant number FA9550-18-1-0056, and the Northrop Grumman NexGen. }
\conflictsofinterest{The authors declare no conflict of interest.}
\reftitle{References}





\end{document}